\documentclass [preprint]{aastex}

\author{M. Freeman\altaffilmark{1}, R. Montez Jr.\altaffilmark{2}, J.~H.\ Kastner\altaffilmark{1},
 B. Balick\altaffilmark{3},
 D.~J. Frew\altaffilmark{4}, D. Jones\altaffilmark{17,18},
 B. Miszalski\altaffilmark{5,6},
 R. Sahai\altaffilmark{7},
 E. Blackman\altaffilmark{8}, Y.-H. Chu\altaffilmark{9},
O. De~Marco\altaffilmark{4},
 A. Frank\altaffilmark{8},
 M.~A. Guerrero\altaffilmark{10},
 J.~A. Lopez\altaffilmark{11},
 A. Zijlstra\altaffilmark{12},
 V. Bujarrabal\altaffilmark{14}, R.L.M. Corradi\altaffilmark{15,16},
J. Nordhaus\altaffilmark{16},
Q.~A. Parker\altaffilmark{4,19},
C. Sandin\altaffilmark{20},
D. Sch\"onberner\altaffilmark{20},
N. Soker\altaffilmark{13},
J.~L. Sokoloski\altaffilmark{21},
M. Steffen\altaffilmark{20},
J.~A. Toal\'a\altaffilmark{10},
T. Ueta\altaffilmark{22},
E. Villaver\altaffilmark{23} }

\altaffiltext{1}{Center for Imaging Science and
Laboratory for Multiwavelength Astrophysics, Rochester Institute of
Technology, 54 Lomb Memorial Drive, Rochester NY 14623 USA}
\altaffiltext{1}{Department of Physics \& Astronomy, Vanderbilt University, Nashville, TN}
\altaffiltext{3}{Dept.\ of Astronomy, University of Washington,
 Seattle, WA}
\altaffiltext{4}{Department of Physics \& Astronomy and Macquarie Research Centre for Astronomy, Astrophysics \& Astrophotonics, Macquarie University, Sydney, NSW 2109, Australia}
\altaffiltext{5}{South African Astronomical Observatory, PO Box 9, Observatory, 7935,
 South Africa}
\altaffiltext{6}{Southern
 African Large Telescope Foundation, PO Box 9, Observatory, 7935,
 South Africa}
\altaffiltext{7}{Jet Propulsion Laboratory, MS 183-900, California Institute of Technology, Pasadena, CA 91109, USA}
\altaffiltext{8}{Dept.\ of Physics \& Astronomy, University of
 Rochester, Rochester, NY}
\altaffiltext{9}{Dept.\ of Astronomy, University of Illinois, Urbana-Champaign, IL}
\altaffiltext{10}{Instituto de Astrof\'{\i}sica de Andaluc\'ia, IAA-CSIC,
Glorieta de la Astronom\'{\i}a s/n, Granada, 18008 Spain}
\altaffiltext{11}{Instituto de Astronom\'ia, Universidad Nacional Autonoma de Mexico, Campus Ensenada, Apdo.\ Postal 22860, Ensenada, B. C., Mexico}
\altaffiltext{12}{School of Physics and Astronomy, University of Manchester, Manchester M13 9PL, UK}
\altaffiltext{13}{Observatorio Astronomico Nacional, Apartado 112,
 E-28803, Alcala de Henares, Spain}
\altaffiltext{14}{Instituto de Astrof{\'{\i}}sica de Canarias, E-38200
La Laguna, Tenerife, Spain}
\altaffiltext{15}{Departamento de Astrof{\'{\i}}sica, Universidad de
 La Laguna, E-38206 La Laguna, Tenerife, Spain}
\altaffiltext{16}{NSF Astronomy and Astrophysics Fellow, Center for Computational Relativity and Gravitation, Rochester Institute of Technology, Rochester, NY 14623}
\altaffiltext{17}{Departamento de F\'isica, Universidad de Atacama, Copayapu 485, Copiap\'o, Chile}
\altaffiltext{18}{European Southern Observatory, Alonso de C\'ordova 3107, Casilla 19001, Santiago, Chile}

\altaffiltext{19}{Australian Astronomical Observatory, PO Box 296, Epping, NSW 2121, Australia}
\altaffiltext{20}{Leibniz Institute for Astrophysics Potsdam (AIP), An der Sternwarte 16, D-14482 Potsdam, Germany}
\altaffiltext{21}{Columbia Astrophysics Laboratory,  Columbia
 University, New York, NY 10027}
\altaffiltext{22}{Dept.\ of Physics \& Astronomy, University of Denver, Denver, CO 80208}
\altaffiltext{23}{Departamento de F\'{\i}sica Te\'orica, Universidad
 Aut\'onoma de Madrid, Cantoblanco
28049  Madrid, Spain}

\title{The {\em Chandra} Planetary Nebula Survey (\chanplans). II.\\X-ray Emission from Compact Planetary Nebulae}
\usepackage [italian,english]{babel}
\usepackage {indentfirst}
\usepackage{multirow}
\usepackage{hyperref}
\usepackage{amsmath,amssymb}
\usepackage{rotating}
\usepackage{morefloats}
\usepackage{longtable}

\usepackage{epstopdf}

\usepackage{natbib}
\newcommand{\chanplans}{{\sc ChanPlaNS}}

\begin{document}
\date{}
\maketitle

\maketitle

\section*{Abstract}

We present results from the most recent set of observations obtained as part of the {\em Chandra} X-ray observatory Planetary Nebula Survey (\chanplans), the first comprehensive X-ray survey of planetary nebulae (PNe) in the solar neighborhood (i.e., within $\sim1.5$ kpc of the Sun). The survey is designed to place constraints on the frequency of appearance and range of X-ray spectral characteristics of X-ray-emitting PN central stars and the evolutionary timescales of wind-shock-heated bubbles within PNe. \chanplans\ began with a combined Cycle 12 and archive {\em Chandra} survey of 35 PNe. \chanplans\ continued via a {\em Chandra} Cycle 14 Large Program which targeted all (24) remaining known compact ($R_{\text{neb}}\lesssim0.4$ pc), young PNe that lie within $\sim1.5$ kpc. Results from these Cycle 14 observations include first-time X-ray detections of hot bubbles within NGC 1501, 3918, 6153, and 6369, and point sources in HbDs 1, NGC 6337, and Sp 1. The addition of the Cycle 14 results brings the overall \chanplans\ diffuse X-ray detection rate to $\sim27\%$ and the point source detection rate to $\sim36\%$. It has become clearer that diffuse X-ray emission is associated with young ($\lesssim5\times10^3$ yr), and likewise compact ($R_{\text{neb}}\lesssim0.15$ pc), PNe with closed structures and high central electron densities ($n_e\gtrsim1000$ cm$^{-3}$), and rarely associated with PNe that show H$_2$ emission and/or pronounced butterfly structures. Hb 5 is one such exception of a PN with a butterfly structure that hosts diffuse X-ray emission. Additionally, of the five new diffuse X-ray detections, two host [WR]-type CSPNe, NGC 1501 and NGC 6369, supporting the hypothesis that PNe with central stars of [WR]-type are likely to display diffuse X-ray emission.

\section{Introduction}

Planetary nebulae (PNe), which represent the near-ends of the lives of intermediate-mass ($\sim$1-8 M$_{\odot}$) stars, afford insight into the late stages of stellar evolution. As an asymptotic giant branch (AGB) star sheds its tenuous envelope, it unveils a hot ($T$ up to $\gtrsim10^5$ K) core that is capable of ionizing the surrounding material and will eventually evolve into a white dwarf (WD). PNe are well known for their wide variety of optical emission line morphologies; this morphological diversity reflects the effects of energetic, nebula-shaping wind collisions. Recently, X-ray observations of PNe have shed light on the processes within nebular interiors, as well as on the properties of their central stars, that directly relate to PN shaping \citep[for a review, see][hereafter Paper I]{Kastner:2012lr}. 

Models describing the structure of PNe \citep{Kwok:1978ab,SchmidtVoigt:1987ab,Marten:1991ab,Villaver:2002ab,Perinotto:2004ab,Toala:2014hb} posit that fast ($v_{\text{w}}\simeq500$-1500 km s$^{-1}$) winds emanating from the pre-WD will collide with the previously ejected AGB envelope, sweeping up the AGB ejecta into a thin shell. These wind collisions result in shocks that heat the fast wind gas to temperatures $>10^6$ K, creating a ``hot bubble" capable of producing soft X-rays ($0.3\sim1$ keV). A binary companion to the central star of the PN (CSPN) is capable of further influencing the PN shape via formation of an accretion disk \citep{Soker:2000ab}, transfer of angular momentum, the generation of a strong magnetic dynamo at the CSPN \citep{Nordhaus:2007ab}, or in the case of a close binary, the formation of jets \citep{Soker:1994ab,Miszalski:2011ab,Corradi:2011ab,Boffin:2012ab,Tocknell:2014ab}. While some CSPNe with hot photospheres ($\gtrsim100$ kK) are capable of emitting soft X-rays, many CSPNe exhibit hard X-rays ($>1$ keV) that cannot be attributed to photospheric emission alone \citep{Montez:2013ab,Montez:2014pt}. Such hard X-rays might be expected, however, when a hot, compact companion accretes material, as suggested for some white dwarfs with hard X-ray emission \citep{Bilikova:2010ab}, or from the corona of a ``rejuvenated" late-type companion of a post-common envelope binary CSPN \citep{Montez:2010ab}. There is also the possibility that a ``hard excess" could result from material infall from a debris disk around the CSPN, similar to the debris disk discovered around the central star of the Helix Nebula and possibly linked to its hard X-ray emission \citep{Su:2007ab}. Alternatively, the hard X-ray emission might emerge from shocks within the stellar wind, via a process similar to that observed in O and B stars \citep{Cassinelli:1994bs}.

The {\em Chandra} Planetary Nebulae Survey ({\sc ChanPlaNS}) is designed to place constraints on the frequency of appearance and range of X-ray spectral characteristics of X-ray-emitting PN central stars and the evolutionary timescales of wind-shock-heated bubbles within PNe. The initial {\sc ChanPlaNS} program consisted of observations of 21 high-excitation PNe selected from among the $\sim120$ known PNe within $\sim1.5$ kpc of Earth, along with another 14 PNe previously observed by {\em Chandra} that fall within this volume (Paper I). The initial results included more than a dozen new detections of X-ray point sources and placed new constraints on the physical conditions, timescales, and frequency of hot bubbles. The survey recently continued with a 670 ks {\em Chandra} Cycle 14 Large Program of an additional 24 compact ($R_{\text{neb}}\lesssim0.4$ pc) PNe selected from among the volume-limited sample. In this paper, we describe the results of the {\sc ChanPlaNS} survey observations of these compact PNe in the context of the overall volume-limited sample of PNe. In two companion papers \citep[in prep]{Montez:2014pt,Montez:2014df}, we detail the characteristics of point-like and diffuse X-ray emission sources within PNe, and further examine the origins and implications of these PN X-ray sources.

\section{Observations and Data Reduction}
	\subsection{Sample: Compact ($R_{\text{neb}} \lesssim 0.4$ pc) planetary nebulae within $\sim1.5$ kpc}
	
The PNe that constitute the sample observed with {\em Chandra} during Cycle 14, listed in Table \ref{table:cycle14}, were primarily selected so as to further probe the onset, evolutionary timescale, and properties of diffuse X-ray sources within PNe. Hence, we selected objects on the basis of their relatively small physical dimensions, as ascertained via optical imaging. More specifically, we chose the 24 Table 1 objects from the compilation of solar-neighborhood PNe in \citet{Frew:2008ab} such that the resulting merged Cycle 12+14+archival sample would constitute a representative, volume-limited ($D \lesssim 1.5$ kpc) sample of PNe with inner bubble or inner shell radii $\lesssim$ 0.4 pc (excluding very low surface brightness PNe, which typically have faint central stars). We estimate that our completeness in surveying compact ($ R_{\text{neb}} \lesssim 0.4$ pc) nebulae within $\sim1.5$ kpc is 90\%, based on the solar neighborhood census of PNe recently compiled by \cite{Frew:2014b}. The chosen nebular radius roughly corresponds to a dynamical PN lifetime of $\lesssim10^4$ yr --- approximately twice the apparent hot bubble dissipation timescale inferred from Cycle 12 plus archival {\em Chandra} data shown in Paper I --- assuming typical PN expansion velocities of $\sim$40 km s$^{-1}$ \citep{Jacob:2013ab}. Basic PN and CSPN data for the sample objects can be found in Table \ref{table:cycle14}.

	\subsection{Observations}

All Cycle 14 PNe were observed with the back-illuminated (BI) CCD of {\em Chandra}'s Advanced CCD Imaging Spectrometer (ACIS). The use of CCDs as X-ray detectors provides determinations of incident photon energies as well as locations, which in turn allows for filtering of images by photon energy. {\em Chandra}/ACIS-S3 has an energy sensitivity of $\sim0.3$-8.0 keV, with a field of view of $\sim8'\times8'$ and pixel size $0.492''$. The BI CCD S3 has greater low-energy ($<1$ keV) sensitivity compared to the front-illuminated ACIS-I array, extending sensitivity down to $\sim0.2$ keV for high soft photon fluxes, albeit with uncertain calibration \citep{Montez:2013ab}. Additionally, use of S3 facilitates subpixel event repositioning (SER) in processing so as to better sample the High Resolution Mirror Assembly core of the point spread function \citep{Li:2004ab}. Observation IDs, dates, and exposure times for the Cycle 14 {\sc ChanPlaNS} observations are listed in Table \ref{table:obs}.

	\subsection{Data reduction}

\begin{figure*}[t]
				\centering$
				\begin{array}{cc}
					\includegraphics[height=40mm]{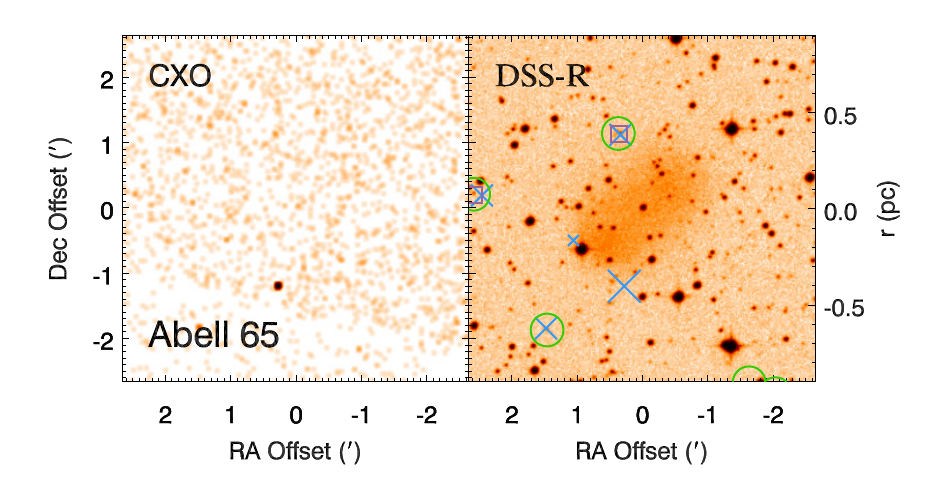}	&\includegraphics[height=40mm]{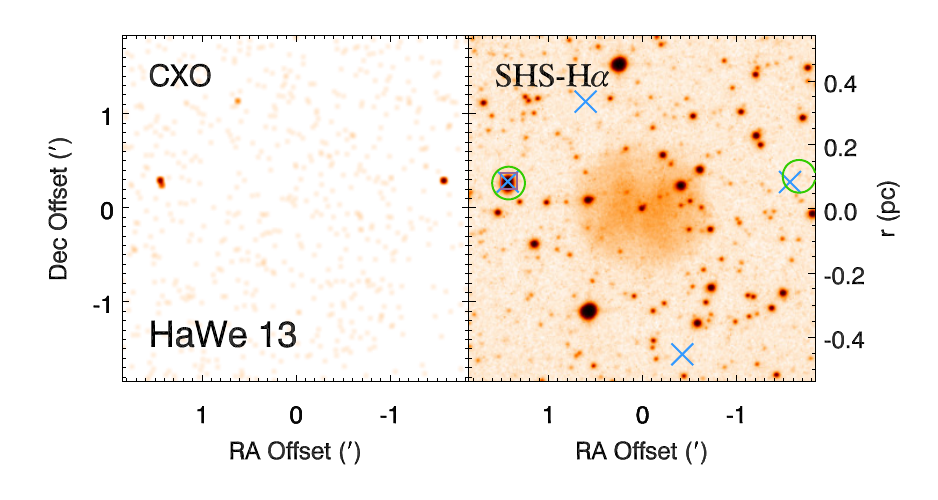}\\
					\includegraphics[height=40mm]{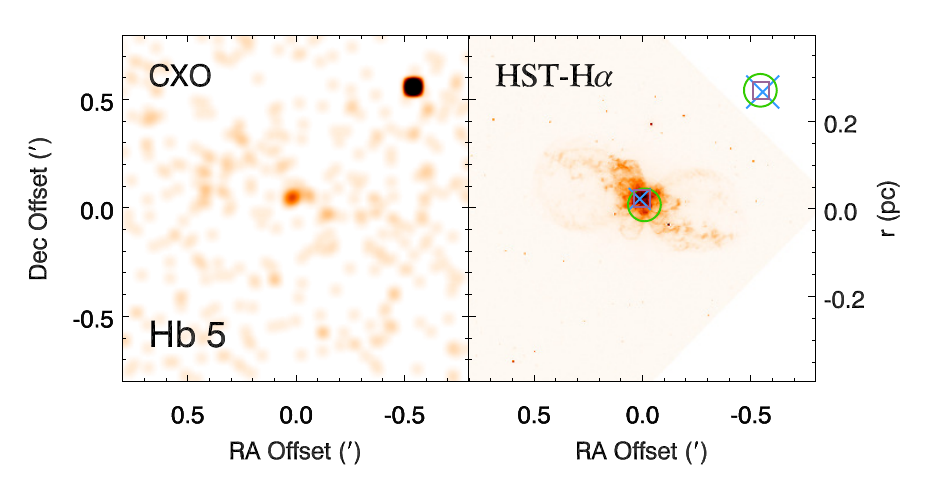}	&\includegraphics[height=40mm]{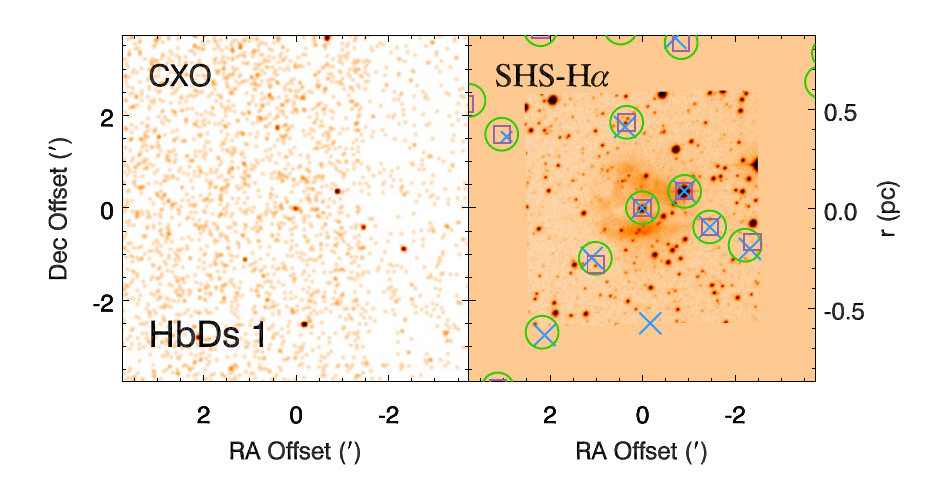}\\
					\includegraphics[height=40mm]{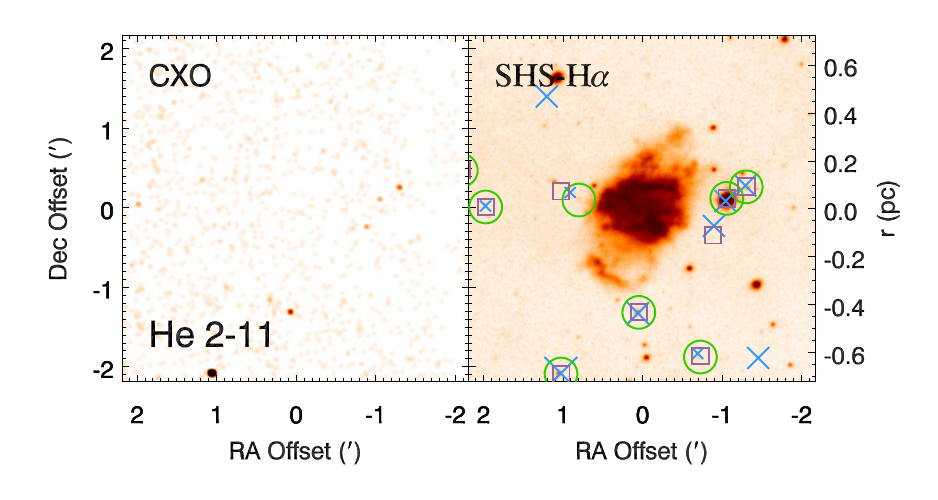}	& \includegraphics[height=40mm]{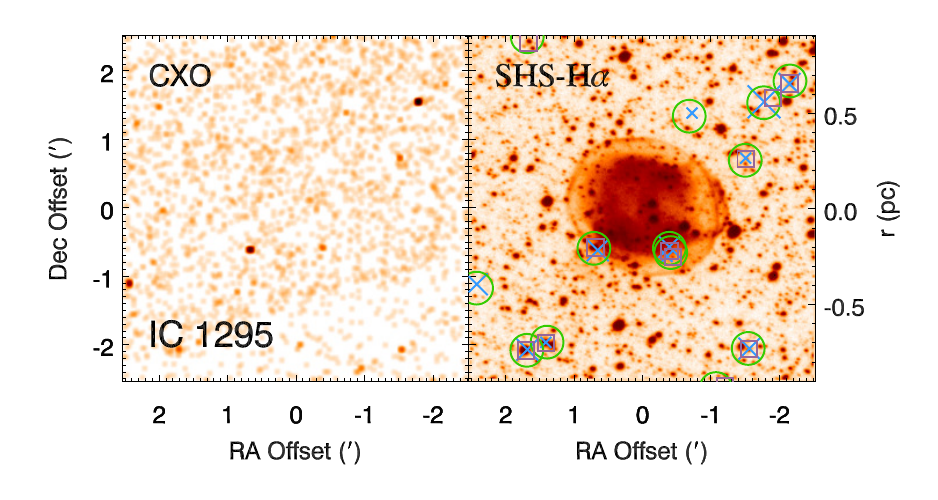}\\
					\includegraphics[height=40mm]{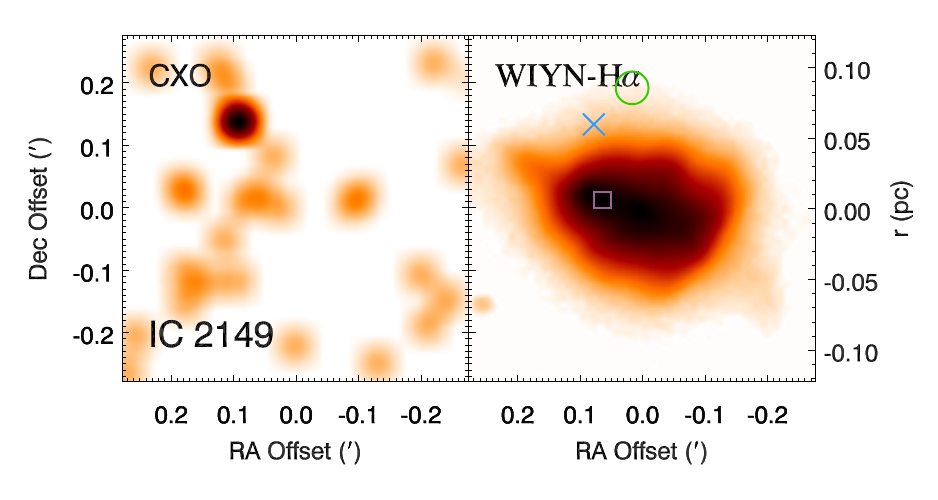}	& \includegraphics[height=40mm]{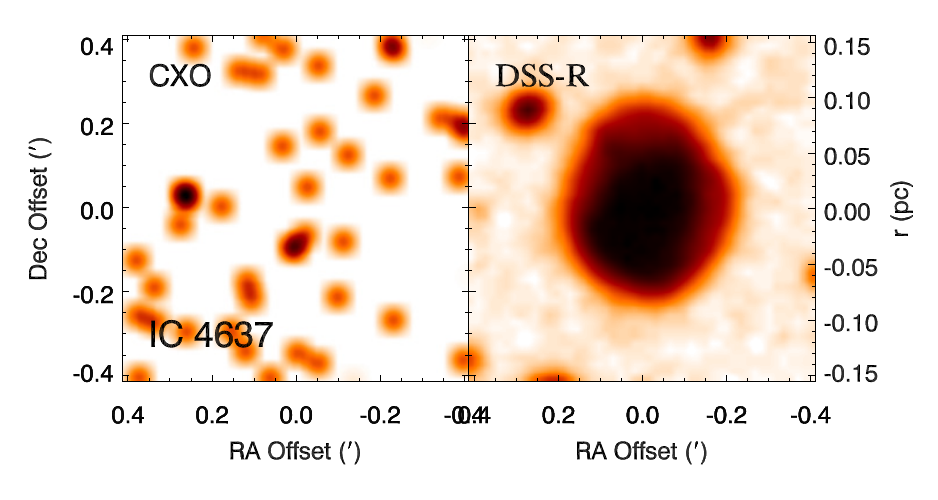}
				\end{array}$
				\caption{\label{fig:images1}\footnotesize{{\sc ChanPlaNS} pipeline output for the Table 1 PNe. Two panels are presented for each PN. The {\em left panel} of each pair shows the {\em Chandra}/ACIS soft-band (0.3-2.0) keV image, smoothed with a Gaussian function with $3''$ FWHM ($1''$ FWHM for M 1-26, or $6''$ FWHM for images larger than $5'$ on a side), centered on the SIMBAD coordinates of the PN (which lies on back-illuminated CCD S3). The {\em right panel} shows an optical image (obtained from {\em HST}, WIYN, the SuperCOSMOS H-alpha Survey, or the DSS, as indicated) overlaid with the positions of detected broad-band (0.3-8.0 keV) X-ray sources (crosses), USNO catalog stars (circles), and 2MASS Point Source Catalog IR sources (squares). The size of the cross is proportional to the number of X-ray photons detected.}}
			\end{figure*}

All data were reduced using CIAO\footnote{\footnotesize{Chandra Interactive Analysis of Observations; http://cxc.harvard.edu/ciao/}} (version 4.5). Data reduction made use of the \chanplans\ pipeline which is detailed in Paper I. Briefly, this pipeline consists of the following steps: reprocessing of the primary and secondary data files and the application of SER using \texttt{chandra\_repro}; detection of sources in the full field of view of each observation \texttt{wavdetect}; calculation of statistics from a $3.5''$ radius region centered on the CSPN, to place constraints on possible X-ray point source emission; and generation of annotated two-panel images displaying a filtered {\em Chandra} X-ray image highlighting the predominately soft X-ray (0.3-2.0 keV) nature of the nebulae and the best available optical (H$\alpha$ or $R$ band) image obtained from the {\em Hubble} Legacy Archive\footnote{\footnotesize{http://archive.stsci.edu/hst}}, the Wisconsin-Indiana-Yale-NOAO (WIYN) 0.9m telescope (see Paper I), the {SuperCOSMOS} H-Alpha Survey \citep[SHS;][]{Parker:2005a,Frew:2014a}, or the Digitized Sky Survey (DSS\footnote{\footnotesize{http://archive.stsci.edu/dss}}), as indicated in the panels (Fig.~\ref{fig:images1}).

\section{Results}

Results from the Cycle 14 {\em Chandra} observations listed in Table \ref{table:obs} are illustrated in Figs.~\ref{fig:images1}-\ref{fig:morph} and are summarized in Tables \ref{table:cycle14} and \ref{table:ptsrcs}. The last column of Table \ref{table:cycle14} states whether the PN was detected and, if so, the type of X-ray emission detected (point-like, diffuse, or both). As in Paper I, our conservative estimate for our sensitivity limits for diffuse and hard (soft) point-like X-ray sources are $\sim10^{30}$ and $\sim10^{29}$ ($\sim10^{31}$) erg s$^{-1}$, respectively, at a distance of 1.5 kpc. We will revisit these sensitivity limits in \cite{Montez:2014pt,Montez:2014df}.

Table \ref{table:stats} (an update of Table 4 from Paper I) lists X-ray detection statistics broken down by PN morphology, H$_2$, detection/nondetection fraction, and known central star binarity for the entire \chanplans\ sample. Figs.~\ref{fig:morph} and \ref{fig:medenergy} represent updates of Figs.~5 and 6, respectively, in Paper I. In Fig.~\ref{fig:morph}, various observed parameters for the sample PNe $-$ optical morphology, H$_2$ detection/nondetection, X-ray detection/nondetection, and X-ray source type $-$ are illustrated in a plot of CSPN effective temperature vs. nebular radius. In Fig.~\ref{fig:medenergy}, we plot median X-ray photon energy against nebular radius. In Fig.~\ref{fig:ne}, which has no counterpart in Paper I, we further illustrated PN X-ray detections and non detections in the form of plots of PN distances and ratios of [\ion{O}{3}] to H$\beta$ fluxes against nebular densities for all (59) \chanplans\ sample PNe.

	\subsection{Cycle 14 detections of X-rays from PNe}

We have detected five of the Table \ref{table:cycle14} PNe in diffuse X-ray emission: Hb 5, NGC 1501, NGC 3918, NGC 6153, and NGC 6369 (Fig.~\ref{fig:diff}). The last four objects are first time detections; Hb 5 was detected previously by {\em XMM-Newton} \citep[see below]{Montez:2009ab}. These detections represent  a $\sim21\%$ (5/24) detection rate of hot bubbles within the Cycle 14 sample. When these detections are combined with the previous {\sc ChanPlaNS} observations, the \chanplans\ detection rate of PNe hot bubbles now stands at $\sim27\%$ (16/59). Two of the newly detected objects, NGC 1501 and NGC 6369, host Wolf-Rayet-type (WR) CSPNe, which brings the \chanplans\ diffuse detection rate of such objects to $100\%$ (5/5). The detection rate of diffuse X-ray emission from [WR]-type CSPNe remains high when including PNe beyond $\sim1.5$ kpc \citep[see][]{Kastner:2008a}. These diffuse X-ray detections follow the previously-established trend that diffuse X-ray PNe generally display compact ($R_{\text{neb}}\lesssim0.15$ pc), elliptical optical morphologies (Fig.~\ref{fig:morph}; Paper I). A notable exception is Hb 5, which has a bipolar morphology.

The Table \ref{table:cycle14} detections of X-ray point sources within three Cycle 14 PNe, HbDs1, NGC 6337, and Sp 1, constitutes a $\sim13\%$ detection rate of CSPNe. These Cycle 14 detections bring the overall {\sc ChanPlaNS} detection rate of CSPNe to $\sim36\%$ (21/59). Two of the three new CSPNe (within NGC 6337 and Sp 1) are known binaries, which brings the overall \chanplans\ detection rate of all CSPN that are known to be binaries (as of writing this paper) to $60\%$ (Table \ref{table:stats}).

	\subsection{PNe displaying diffuse X-ray emission}
	
The diffuse X-ray emitting PNe detected in the Cycle 14 sample are displayed in Fig.~\ref{fig:diff} and are described briefly below. As is evident in Figs.~\ref{fig:morph} and \ref{fig:medenergy} and Table \ref{table:stats}, these detections reinforce previous \chanplans\ results indicating that PNe displaying diffuse X-rays are predominantly elliptical in morphology with $R_{\text{neb}}\lesssim0.15$ pc; furthermore, diffuse X-ray PNe generally lack detections of near-IR H$_2$ emission. In a forthcoming paper \citep[in prep]{Montez:2014df}, we present a global analysis of the \chanplans\ diffuse X-ray sources. Here, we describe the properties of the diffuse X-ray PNe detected during Cycle 14 observations.

		\subsubsection{Classical hot bubbles}

NGC 3918 appears to host multiple shells, and is similar in optical and X-ray morphology to multiple-shell, diffuse X-ray PNe such as NGC 6543 (Paper I). Three-dimensional models of the nebula vary in their interpretations of its morphology. \cite{Clegg:1987ab} modeled NGC 3918 as a biconical nebula within an optically thin sphere, while a spindle-like structure was favored by both \cite{Corradi:1999ab} and \cite{Ercolano:2003ab}. Like NGC 1501, the shell of the nebula shows enhanced [\ion{O}{3}]/H$\alpha$ emission resulting from the inner shell expanding into the outer shell \citep{Guerrero:2013ab}, while the inner shell confines the diffuse X-ray emission.

NGC 6153 is also an elliptical nebula that hosts multiple shells, and is thought to be the product of a high mass (4.5 M$_{\odot}$) progenitor \citep{Pottasch:2010ab}. Often compared to NGC 7009, NGC 6153 is indeed morphologically similar, both in the optical and X-ray; however NGC 6153 lacks the ansae found in NGC 7009. The diffuse X-ray emission is confined to the inner shell, as is the case with NGC 3918 and similar multiple-shell, diffuse X-ray PNe (Paper I).

		\subsubsection{PNe with [WR]-type CSPNe}
		
Both NGC 1501 and NGC 6369 harbor [WR]-type central stars and diffuse X-ray emission. NGC 1501 is an elliptical nebula with a [WR]-type central star within a complex, filamentary internal structure. The central star has been reclassified many times between [WO] and [WC] type. Based on emission lines detected in the optical spectrum, the CSPN appears to be [WO4] \citep{Crowther:1998ab,Ercolano:2004ab}; however, the star also displays an infrared excess from $J$ through all {\em Spitzer} IRAC bands, which is indicative of circumstellar dust, as is common for [WC]-type stars \citep{Bilikova:2012ab}. NGC 1501 shows enhanced [\ion{O}{3}]/H$\alpha$ emission along the surface of the nebular shell, which is evidence for shocks along the skin \citep{Guerrero:2013ab}. Similarly, the X-ray emission appears to trace shocks inside the primary nebular shell.

In contrast to the elliptical shell and filamentary internal structure of NGC 1501, NGC 6369 appears to have a spherical main nebular shell described as a barrel, with bipolar outflows \citep{Ramos-Larios:2012ab}. This structure is similar to that of NGC 40 \citep{Montez:2005ab}. Where [\ion{O}{3}] and H$\alpha$ are dim in NGC 6369, the nebula is bright in [N II], displaying a ring of emission along the inner shell. The bright optical shell is filled with H$_2$ and surrounded by a photodissociation region \citep{Ramos-Larios:2013ab}, and the shell encloses the diffuse X-ray emission around the [WO]-type central star \citep{Pena:2013ab}. The correspondence of the X-ray emission and the nebular shell, and lack of emission from the blowouts, suggests that emission is generated by the strong winds of the CSPN colliding with the progenitor red giant wind, as is the case for NGC 40 \citep{Montez:2005ab}.

		\subsubsection{The bipolar planetary nebula, Hb 5}

Hb 5 is a bipolar nebula with large lobes ($\sim27''$) extending east and west and a tight waist with a bright core that contains secondary expanding lobes. The smaller lobes ($\lesssim12''$) may result from an interaction between a fast, low-density wind and an inhomogeneous, high density circumstellar shell \citep{Lopez:2012ab}. Diffuse X-ray emission was previously detected serendipitously by {\em XMM-Newton} \citep{Montez:2009ab}. The emission in the {\em XMM} images appeared coincident with the peak optical emission around the core and possibly along the H$\alpha$ emission in the lobes, although {\em XMM}'s resolution was insufficient to determine if the X-ray emission is extended.

The recent {\em Chandra} observations also detected a bright, compact X-ray core, coincident with the peak optical emission. There is no apparent X-ray point source from the CSPNe contributing to the core emission. Additionally, X-ray emission was tentatively detected along the tip of the eastern lobe, which appears blueshifted, while no apparent emission is seen from the western, redshifted lobe which may be attributed to extinction by intervening nebular material \citep{Lopez:2012ab}.

	\subsection{PNe displaying point-like X-ray emission at central stars}
	
The three Cycle 14 X-ray point source detections (HbDs1, NGC 6337, Sp 1) continue the trend, noted in Paper I, that soft ($\sim0.6-1.0$ keV) X-ray sources likely indicate the presence of CSPN photospheric emission and/or wind shocks, while hard ($\gtrsim1.0$ keV) sources appear related to the presence of binary companions to CSPNe \citep[Fig.~\ref{fig:medenergy};][in prep]{Montez:2014pt}. Below are brief summaries of each of the Table \ref{table:cycle14} PNe with newly detected point-like X-ray emission.

HbDs1 is an irregularly elliptical nebula around the hot blue star LSS 1362 \citep[$T_{\text{eff}}=116$ kK;][]{Herald:2011ab}. The nebula has enhanced H$\alpha$ emission along its northern and southern limbs. The magnetic field properties of the CSPN have been debated. A kG field was inferred by \cite{Jordan:2005ab,Jordan:2012ab} from spectropolarimetric data taken with the ESO Very Large Telescope; however, more recent FORS2 observations showed no evidence of such a strong field \citep[with upper limits of $B<600$ G;][]{Leone:2011ab,Bagnulo:2012ab}. The detected X-ray point source has a median energy of 0.65 keV, which may indicate self-shocking within the stellar winds previously detected at the CSPN \citep{Herald:2011ab,Guerrero:2013bc}, possibly combined with photospheric emission \citep[see][]{Montez:2013ab}.

NGC 6337 is seen as a ring nebula with radial filaments. This PN harbors a nearly face-on ($i \lesssim 15^{\circ}$) close binary (0.17 d period) central star system \citep{Hillwig:2010ab}. It is likely that the nebula is viewed nearly pole-on as well, suggesting a bipolar structure with a narrow waist around the binary \citep{Garcia-Diaz:2009ab}. The close binary is most likely the product of a common envelope phase \citep{DeMarco:2011ab}, leaving behind a pre-WD and a cool main-sequence star \citep{Hillwig:2010ab}. The X-ray point source associated with this binary has a median energy of 1.27 keV, the hardest emission detected from a point source in the \chanplans\ sample thus far. This hard X-ray detection provides strong support for a model in which the X-rays originate from a cool, coronally active companion (see \S 4).

Sp 1 appears similar to NGC 6337; this PN is also a diffuse ring on the plane of sky and harbors a close binary (2.9 d period) at its center \citep{Bond:1990ab,Jones:2012ab,Bodman:2012ab}. Spatiokinematical modeling of the nebular shell shows that the nebula is a nearly pole-on hourglass shape \citep{Jones:2012ab}. This is borne out by modeling of the CSPN by \cite{Bodman:2012ab}, which indicates a G0 or F8 main sequence companion with an orbital inclination of $10^{\circ}<i<20^{\circ}$, consistent with the generation of a near pole-on nebula. UV spectra of the CSPN show evidence for stellar winds \citep{Patriarchi:1991ab}, which could be at best partly responsible for the X-ray point source at the CSPN. However, the median energy for the X-rays from the Sp 1 CSPN is 0.93 keV, making it one of the harder X-ray sources in the {\sc ChanPlaNS} sample, and suggesting that the companion dominates the detected emission.

	\subsection{X-ray nondetections}

Table \ref{table:ptsrcs} lists the point source X-ray characteristics (net counts, count rates, median photon energies, and photon energy ranges) of each PN in the Cycle 14 sample. To obtain these results, the 0.3-3.0 keV background counts were extracted from a source-free region near the central star. These background counts were scaled to the point source extraction radius of $3.5''$, and then subtracted from the counts of the same extraction region centered on the CSPN, to obtain the net source counts (or source count upper limits). Sources with less than 6 net counts are classified as undetected. Two sources lie below this background-subtracted limit, Abell 65 and NGC 6026, but could be considered as tentative detections; there are $\sim4$ (source+background) counts found near their CSPNe, both of which have companion stars. There are a few sources above the net count limit that we consider to be doubtful as point source detections: Hb 5, NGC 3918 and NGC 6369. Contamination from the diffuse X-ray emission in NGC 3918 makes it difficult to determine the contribution from a possible point source at the CSPN in the circular source region. NGC 6369 is a luminous, albeit reddened, PN that lies near the limit of our volume-limited sample, at a distance of 1.55 kpc \citep{Monteiro:2004ab}. Additionally, the diffuse X-ray emission of NGC 6369 is not concentrated at the CSPNe, which makes it unlikely to contaminate a point source detection. Finally, as noted in \S 3.2, Hb 5 appears to have no point-like emission within the CSPN extraction region.

Several PNe that were undetected appear to fit within morphology trends typical of previous \chanplans\ point source and diffuse X-ray detections. IC 5148/50 is undetected, yet it has a morphology similar to Sp 1, suggesting that it may be a pole-on hourglass and, hence, could be the product of a binary system. In optical images, NGC 6894 has the multiple-shell structure of a hot bubble from which we would expect to see diffuse X-ray emission, though it is important to note that the central star is faint and on the WD cooling track. Similarly, NGC 7354 hosts multiple elliptical shells and ansae, similar to NGC 3918 and NGC 6153; however no X-ray photons were detected from the innermost bubble of NGC 7354. These nondetections appear to place stringent limits on the timescales for X-ray emitting hot bubbles within PNe (\S 4.2) $-$ although we note that NGC 7354 is quite heavily reddened and, hence, any soft X-rays from its interior might suffer considerable absorption.

The objects with no detectable X-ray emission have varying morphologies (round, elliptical, bipolar), with at least half exhibiting inherent asymmetries. IC 4637 is an example of a PNe that falls within the range of nebular properties for the diffuse X-ray detections (Paper I), but was not detected. Specifically, IC 4637 has a radius of 0.05 pc and elliptical morphology \citep{Frew:2008ab}. Additionally, IC 2149 is a young, compact bipolar nebula with an elliptical central shell and may be the product of a low mass progenitor and/or binary evolution \citep{Vazquez:2002ab}. Upper limit estimates for X-ray luminosity $L_X$ for each of these sources \citep[in prep]{Montez:2014df} will help place constraints on the sizes, plasma densities, and morphologies of X-ray emitting nebulae.

\section{Discussion}

The sample of 59 PNe within $\sim1.5$ kpc observed to date by {\em Chandra} constitutes a representative set of X-ray observations of mostly compact and, hence, predominately relatively young PNe within the solar neighborhood. Adding the 24 PNe observed in Cycle 14 to the initial sample (see Paper I) reduces the overall PN X-ray detection rate from $\sim70\%$ to $\sim54\%$ (32/59). This decrease most likely reflects the composition of the initial {\sc ChanPlaNS} sample; these objects were chosen based on the presence of bright excitation lines indicative of hot CSPNe ($T_{\text{eff}}\gtrsim10^5$ K) and/or structure that resulted from rapid evolution (Paper I), while the Cycle 14 objects were chosen to complete the X-ray census of compact ($R_{\text{neb}}<0.4$ pc) PNe within $\sim1.5$ kpc. Nevertheless, the Cycle 14 results underscore trends apparent in the X-ray properties of PNe, as we discuss in the following subsections.

	\subsection{Diffuse X-ray emission from [WR]-type CSPNe}
	
The CSPNe that are H-deficient and display spectroscopic evidence for fast stellar winds and high mass-loss rates \citep[up to $10^{-6}$ M$_{\odot}$ yr$^{-1}$;][]{Crowther:1998ab,Depew:2011ab} are classified as [WR]-type because of their similarity, as a class, to luminous WR stars. The primary difference between the two classes is that the [WR]-type stars descend from intermediate mass (1-8 M$_{\odot}$) stars, whereas ``classical" WR stars are the descendants of massive ($>25$-30 M$_{\odot}$) stars \citep{Crowther:2006ab}. In total there are 5 known [WR]-type CSPNe in the {\sc ChanPlaNS} sample: NGC 1501 and NGC 6369 from Cycle 14 (Table \ref{table:cycle14}), as well as NGC 40, NGC 2371, and BD +30$^{\circ}$3639 from Cycle 12 and archival observations (Paper I). All 5 of these [WR]-type CSPN objects display diffuse X-ray emission. The only consistent morphological similarity among all of the [WR] type CSPNe is that their main bubbles appear elliptical \citep{Frew:2008ab}. In each case, the diffuse X-ray emission is enclosed by the central optical bubble, although the morphology of the X-ray emission varies considerably from object to object.

\begin{description}
	\item{\bf BD +30$^{\circ}$3639:} The unusually bright, diffuse emission fills the central bubble, perhaps as a consequence of its very young age \citep[$\sim10^3$ yr;][]{Frew:2008ab}.
	\item{\bf NGC 40:} The emission appears within a partial ring that traces the inner edge of the bright optical rim of the nebula \citep{Montez:2005ab}.
	\item{\bf NGC 2371:} The diffuse emission is contained within the bright nebular rim but does not seem to follow any significant optical feature, and is highly asymmetric.
	\item{\bf NGC 1501:} The X-ray emission traces the rim of the nebula, echoing the limb-brightened morphology of WR star wind-blown bubble S308 \citep{Toala:2012fk}.
	\item{\bf NGC 6369:} The diffuse emission is asymmetrical and strongest along the northern limb of the central bubble.
\end{description}

The X-ray results for most of the above PNe, with the possible exception of NGC 2371, reveal some potential similarities between PNe with [WR]-type CSPNe and the wind-blown bubbles of massive WR stars. The X-ray limb-brightening in massive WR star wind-blown bubbles, such as S308, is the result of the hot WR wind cooling as it comes in contact with the the previously ejected red (or yellow) supergiant or luminous blue variable material \citep{Garcia-Segura:1996b,Garcia-Segura:1996a,Freyer:2003a,Toala:2011a,Toala:2012fk,Dwarkadas:2013a}. A similar process occurs in PNe, wherein the X-ray emission is contained within the ejected AGB envelope and does not come in direct contact with the ISM. The question remains: {\em is the diffuse X-ray emission from [WR] CSPNe limb-brightened or not?} Given the photon-starved nature of our detections, we will require Poisson-based models for comparison with the observations to adequately answer this question.

	\subsection{Diffuse X-ray emission and PN size, age, and electron density}
	
The diffuse X-ray emission from hot bubbles in PNe is evidently confined to early phases in PN evolution. In Paper I, we pointed out that the central bubbles of PNe with diffuse emission tend to have elliptical, nested-shell morphologies and radii $\lesssim0.15$ pc (Fig.~\ref{fig:morph} and Tables \ref{table:cycle14} and \ref{table:stats}). This radius limit corresponds to dynamical ages $\lesssim5\times10^3$ yr, roughly a quarter of the $(21\pm5)\times10^3$ yr mean visibility time of PNe \citep{Jacob:2013ab}. Our Cycle 14 detections and nondetections thereby reinforce the inference, made in Paper I, that diffuse X-ray emission is restricted to the most compact nebulae. We note that the correlation of diffuse X-rays with elliptical morphology is ambiguous (Table \ref{table:stats}), but it remains the case that PNe with closed structures (whether bubbles or lobes) are more likely to be detected as diffuse X-ray sources.

From Fig.~\ref{fig:ne} we see, furthermore, that with the exception of NGC 2371, all \chanplans\ PNe with diffuse X-ray emission have nebular densities $n_e\gtrsim1000$ cm$^{-3}$, as determined via their H$\alpha$ line luminosities and nebular radii which are in turn determined from the H$\alpha$ surface brightness vs.~radius relation \citep{Frew:2008ab,Frew:2014b}. In contrast, the diffuse X-ray detection rate is essentially independent of distance and nebular excitation. Since PN density and radius are correlated, the remarkably sharp $n_e$ boundary between diffuse X-ray detections and nondetections in Fig.~\ref{fig:ne} further underscores the notion of a limit on the timescale for energetic (nebula-shaping) wind interactions in PNe. Moreover Fig.~\ref{fig:ne} suggests that, from spectroscopy of density-sensitive lines (e.g. [\ion{O}{2}], [\ion{S}{2}]), it is possible to ascertain whether a PN might currently be in a wind-collision phase, when we would expect to see diffuse X-ray emission. Nevertheless, the duration of the wind-interaction phase is a complex problem that must account for (1) luminosity of the central star (e.g.~Hb 5 high, NGC 5148 low), (2) expected hydrogen column density, and (3) spectral type of the central star (i.e. whether [WR] or not).

This density limit may explain the X-ray nondetection of nebulae like NGC 6894 and NGC 6804 (Paper I), which appear morphologically similar to known diffuse X-ray emitters. NGC 6894 has a nebular radius of 0.17 pc and NGC 6804 has a nebular radius of 0.19 pc, both of which have a density $\log{n_e}\sim2.7$. Thus, evidently, both NGC 6894 and NGC 6804 lie close to, but just below, the $\log{n_e}$ X-ray detection threshold. It is important to keep in mind that electron density is not a good indicator of PN evolutionary state. Fast evolving stars achieve low luminosities at still quite high nebular densities, but remain undetected in diffuse X-rays \citep{Steffen:2008a}. Upper limit estimates and observations of additional nebulae similar to these $-$ with $n_e\gtrsim1000$ cm$^{-3}$ and closed optical morphologies (multiple shells, well-defined inner bubbles) $-$ will allow us to answer questions about diffuse emission such as: {\em What is the lifetime of the hot bubble shaping phase? How does it relate to the mass of the CSPN progenitor? What is the relationship between diffuse X-ray emission and the final mass of the core, i.e.~the WD remnant?} We aim to revisit the question of diffuse X-ray dependence on age in \cite{Montez:2014df} \citep[also see][]{Schoenberner:2014a}.

	\subsection{Binary detections/nondetections}
	
Six additional known binary CSPNe were observed in Cycle 14, all of which are close binaries ($p<3$ d). Two of these six, NGC 6337 and Sp 1, had detectable point source X-ray emission. Both PNe appear to have nearly pole-on hourglass structures. The median energy for both point sources is relatively hard ($\gtrsim1.0$ keV), which implies that the CSPN X-ray emission is due to late-type companions that display elevated levels of coronal activity. These binary, X-ray-luminous CSPNe are therefore similar to the CSPNe in DS 1, HFG 1, and LoTr 5, which are thought to be systems in which mass transfer from PN progenitor to late-type companion has spun up the companion \citep{Jeffries:1996is,Soker:2002jk,Montez:2010ab}; alternatively, in many or most of these cases of X-ray-emitting binary CSPNe, it is possible the coronal activity of the companion is due to tidal locking that has resulted in rapid, synchronous rotation \cite[a possibility we further explore in][]{Montez:2014pt}. Both NGC 6337 and Sp 1 represent the two most compact (and therefore youngest) PNe harboring X-ray point sources at known binary CSPNe in the \chanplans\ sample (Fig.~\ref{fig:medenergy}). The periods of the close binaries within Abell 65 \citep[1.00 d;][in prep]{Bond:1990ab,Shimansky:2009ab,Hillwig:2014ip}, He 2-11 \citep[0.61 d;][]{Jones:2014ab}, and NGC 6026 \citep[0.53 d;][]{Hillwig:2010ab} lie between those of NGC 6337 \citep[$0.17$ d;][]{Hillwig:2010ab} and Sp 1 \citep[$2.91$ d;][]{Bond:1990ab}, but no point source X-ray emission was detected from the former three PNe. This leaves us with unanswered questions:  {\em What is the relationship between binary period/separation and X-ray point source emission? What are the characteristics of secondaries within binaries that do exhibit X-ray emission?}

IC 5148/50 is a potentially significant nondetection, given its morphological similarity to Sp 1. Based on imagery alone, Sp 1 appears to have a spherical morphology, but spatiokinematcial modeling reveals this is a projection effect and its intrinsic is that of a pole-on hourglass \citep{Jones:2012ab}. Such bipolar structures are likely to be the products of binary interactions \citep[see, e.g.,][and references therein]{Balick:2002ab}. The nondetection of IC 5148/50 may indicate that, while it is morphologically similar to Sp 1, structurally it may be quite different. From H$\alpha$, [\ion{N}{2}], and [\ion{O}{3}] observations \citep{Hua:1998ab} it is clear that IC 5148/50 hosts not only internal elliptical structure but also a helical structure. Observations of H$_2$  \citep{Kastner:1996ab} and/or spatially resolved, high-resolution emission line spectroscopy (for use in spatiokinematical modeling) are necessary to determine whether or not IC 5148/50 has inherently bipolar vs.~elliptical. The CSPN of IC 5148/50 also has no infrared excess indicative of an unresolved cool companion \citep{DeMarco:2013ab,Douchin:2014a}.

It is possible that many other, similarly X-ray-luminous CSPN have gone undetected in our sample of PNe. From Fig.~\ref{fig:ne} we see that the fraction of PNe with detected point-like X-ray emission is strongly dependent on distance. All sample PNe within $\lesssim0.6$ kpc have point-like emission, whereas beyond 1.3 kpc the detection fraction drops to $\sim35$\%. This drop in detection fraction is likely the result of increased interstellar extinction obscuring the softer X-ray point sources, and furthermore implies that we lack the sensitivity necessary to detect even the (harder) binary CSPNe at distances much beyond $\sim1$ kpc. Further discussion concerning the implications of the distance dependence of the X-ray detection rate of CSPNe can be found in \cite{Montez:2014pt}.

	\subsection{Diffuse X-ray emission from jet-wind interactions during common envelope ejection?}
	
Several elliptical PNe display diffuse X-ray emission morphologies wherein the elongated, diffuse emission appears to include peaks on two opposite locations along the long axis of the elliptical rim. Examples are NGC 3242, NGC 6543, NGC 7009, NGC 7662 (Paper I) and NGC 3918 (Fig.~\ref{fig:images1}). One possible interpretation of such an X-ray morphology, with far-reaching implications, is that the diffuse emission emanates from gas heated by shocks as low mass jets with velocities of a few 100 km s$^{-1}$ passed through material previously ejected in a common envelope interaction \citep{Akashi:2008a}. Such post-common envelope jets would most likely be launched by an accretion disk around a main sequence companion to the PN progenitor star during the progenitor's post-AGB phase. Indeed, it is possible that such jets are the ``last gasps'' of outflows launched during the final stages of common envelope evolution \citep{Tocknell:2014ab}. If so, then the diffuse X-ray emission from elliptical PNe might hint at the operation of jets in the final removal of common envelopes formed by evolved, close binary systems. The X-ray emission morphologies of elliptical PNe, and their potential implications for common envelope models of PN formation, will be further explored in subsequence papers in the \chanplans\ series  \cite[e.g.,][in prep]{Montez:2014df}.

\section{Summary}

We present results from the most recent set of observations obtained by the {\em Chandra} Planetary Nebula Survey ({\sc ChanPlaNS}), the first comprehensive X-ray survey of PNe in the solar neighborhood (within $\sim1.5$ kpc). The survey is designed to place constraints on the frequency of appearance and range of X-ray spectral characteristics of X-ray-emitting PN central stars and the evolutionary timescales of wind-shock-heated bubbles within PNe. {\sc ChanPlaNS} began with a combined Cycle 12 and archive {\em Chandra} survey of 35 PNe. {\sc ChanPlaNS} continued via a {\em Chandra} Cycle 14 Large Program which targeted all (24) remaining known compact ($R_{\text{neb}}\lesssim0.4$ pc), young PNe that lie within $\sim1.5$ kpc. The highlights of the \chanplans\ Cycle 14 results include the following:
\begin{itemize}
	\item The overall X-ray detection rate for relatively compact ($\lesssim0.4$ pc) PNe within $\sim1.5$ kpc observed thus far by {\em Chandra} is $\sim54\%$.

	\item Nearly $40\%$ of the \chanplans\ sample PNe host X-ray-luminous point sources at their CSPNe. This includes three new detections of CSPNe X-ray sources within the Cycle 14 sample PNe.
	
	\item Roughly 60\% of the known binaries within the \chanplans\ sample display point-like X-ray emission, including first-time Cycle 14 detections of X-rays from the CSPNe of NGC 6337 and Sp 1.

	\item All PNe with [WR]-type CSPN within the \chanplans\ sample observed thus far (BD $+30^{\circ}3639$, NGC 40, NGC 2371, NGC 1501, NGC 6369) display diffuse X-ray emission. With the exception of NGC 2371, all of these PNe have elliptical morphologies. In each case, the diffuse X-ray emission resembles the limb-brightened morphology of wind-blown bubbles blown by massive WR stars.
	
	\item The addition of the Cycle 14 results brings the overall {\sc ChanPlaNS} diffuse X-ray detection rate to $\sim27\%$.
	
	\item It has become clearer that diffuse X-ray emission is associated with young ($\lesssim5\times10^3$ yr) PNe with compact ($R_{\text{neb}} \lesssim 0.15$ pc), closed structures and high central electron densities ($n_e\gtrsim1000$ cm$^{-3}$). Typically, such nebulae display nested-shell elliptical morphologies and rarely show H$_2$ emission and/or pronounced butterfly structures. Hb 5 is a notable exception.
	
	\item The \chanplans\ detection rate of diffuse X-ray emission from PNe within $\sim1.5$ kpc appears largely independent of distance and excitation, if the latter is measured via the flux ratio of [\ion{O}{3}] and H$\beta$ lines.
	
	\item In contrast, the \chanplans\ detection rate of point-like X-ray emission from PNe within $\sim1.5$ kpc appears strongly dependent on distance, reflecting the relatively low intrinsic X-ray luminosities of CSPNe combined with the effects of interstellar extinction.
	
\end{itemize}
We note that in addition to the need to acquire data for a larger sample of PNe $-$ particularly high-$n_e$ PNe and those harboring known binary central stars $-$ we must carry out estimations of upper limits on $L_X$ in order to reaffirm our claims and to properly answer our open questions. Further analysis of these and future \chanplans\ data and results describing both point-like X-ray emission from CSPNe \citep[in prep]{Montez:2014pt} and diffuse X-ray emission from PNe \citep[in prep]{Montez:2014df} will continue to refine models of PN shaping mechanisms, X-ray emission timescales, and the role of binarity in PN formation and evolution.

{\em We thank the anonymous referee for helpful comments. This research was supported via award number GO3--14019A to RIT issued by the Chandra X-ray Observatory Center, which is operated by the Smithsonian Astrophysical Observatory for and on behalf of NASA under contract NAS8Ð03060. Jes\'us A. Toal\'a and Mart\'in A. Guerrero are supported by the Spanish MICINN grant AYA 2011-29754-C03-02 co-funded with FEDER funds. The Digitized Sky Surveys were produced at STScI under U.S. Government Grant NAG W-2166. This research has also made use of the SIMBAD database, operated at CDS, Strasbourg, France.}

 \bigskip

 \bibliography{chanplans_bib2}

\begin{deluxetable}{l c c c c c c c c c c}
\tabletypesize{\footnotesize}
\tablecolumns{11} 
\tablewidth{0pt} 
\tablecaption{{\sc Planetary Nebulae within 1.5 kpc\tablenotemark{a} Observed by {\em Chandra}}}
\tablehead{\colhead{Name}			&
	\colhead{PN G}					&
	\colhead{morph.\tablenotemark{b}}	&
	\colhead{$D$}					&
	\colhead{$R_{\text{neb}}$}		&
	\colhead{age}					&
	\colhead{$T_{*}$}				&
	\colhead{sp. type}				&
	\colhead{comp.\tablenotemark{c}}	&
	\colhead{H$_2$\tablenotemark{d}}	&
	\colhead{X-rays\tablenotemark{e}}	\\
	\colhead{}						&
	\colhead{}						&
	\colhead{(F08/SMV11)}					&
	\colhead{(kpc)}					&
	\colhead{(pc)}					&
	\colhead{($10^3$ yr)}			&
	\colhead{(kK)}					&			
	\colhead{}						&
	\colhead{}						&
	\colhead{}						&
	\colhead{}						}
\startdata
					Abell 65      	&017.3$-$21.9     	&Eafm:/Ls			&1.17     &0.30       	&17.5\tablenotemark{f}		&114\tablenotemark{g}		&O(H)   	&(Y)		&...	&N\\
   					HaWe 13      	&034.1$-$10.5     	&Efp?/Esip		&1.01     &0.18      	&...       	&68        	&hgO(H) 	&...        	&...	&N\\
					Hb 5      		&359.3$-$00.9     	&Bps/Bcbmph		&1.70     &0.13      	&1.5\tablenotemark{h}      	&172     	&...           	&...        	&Y	&D\\
					HbDs 1      	&273.6$+$06.1		&Er?	/Is			&0.80     &0.29      	&...      	&119\tablenotemark{i}       	&O(H) 	&...        	&...	&P\\
					He 2-11      	&259.1$+$00.9		&Ebps/Bcbsip		&1.14     &0.24      	&0.7\tablenotemark{j}       	&108\tablenotemark{j}           &...   		&K5V\tablenotemark{j}        	&...	&N\\
					IC 1295      	&025.4$-$04.7		&Efm:/Es(b/ti)		&1.23     &0.30      	&11       	&98        	&hgO(H) 	&...        	&...	&N\\
					IC 2149      	&166.1$+$10.4		&E/Bsh			&1.52     &0.04        	&2       	&42         	&Of(H) 	&...        	&N	&N\\
					IC 4637      	&345.4$+$00.1		&Eam/Estp		&1.30     &0.05        	&2       	&50          	&O(H)  	&(Y?)      	&...	&N\\
					IC 5148/50      	&002.7$-$52.4		&Rm	/Rsiph		&0.85     &0.27        	&5      	&110       	&hgO(H)	&...        	&...	&N\\
					M 1-26      		&358.9$-$00.7		&R/Mcbsih		&1.20     &0.02        	&1       	&33         	&Of(H) 	&...        	&...	&N\\
					M 1-41      		&006.7$-$02.2		&Bs/(I/B)			&1.47     &0.15      	&...      	&187         &...           	&...        	&Y	&N\\
					NGC 1501      	&144.5$+$06.5		&Es/Esph			&0.72     &0.09        	&2      	&135         &[WC4] 	&...        	&...	&D\\
					NGC 2899      	&277.1$-$03.8		&Baps/Bosbp		&1.37     &0.37       	&14      	&215\tablenotemark{k}        	&...    	&F5V:        &Y	&N\\
					NGC 3918      	&294.6$+$04.7		&Ems(h)/Lsbairh	&1.84     &0.08        	&3      	&150         &O(H)?  	&...        	&...	&D\\
					NGC 6026      	&341.6$+$13.7		&Ef/Ish			&1.31     &0.16        	&6       	&35           &O7(H)     &WD?       	&...	&N\\
					NGC 6072      	&342.1$+$10.8		&Ba/Mcot			&1.39     &0.23       	&23      	&140        	&...           	&...      	&Y	&N\\
					NGC 6153      	&341.8$+$05.4		&Es/Esah			&1.10     &0.07        	&4      	&109         &Of(H)?  	&...      	&N	&D\\
					NGC 6337      	&349.3$-$01.1		&Epr	/Rsarh		&0.86     &0.10       	&12      	&105         &...           	&(Y)       &...	&P\\
					NGC 6369      	&002.4$+$05.8		&Ebpr(h:)/Mcst		&1.55     &0.12        	&3       	&66         	&[WO3]	&...        	&Y	&D\\
					NGC 6894      	&069.4$-$02.6		&Emr/Eh			&1.31     &0.17        	&4      	&100         &...           	&...        	&...	&N\\
					NGC 7076      	&101.8$+$08.7		&Ea/Esh			&1.47     &0.20        	&5       	&80           	&...           	&...        	&...	&N\\
					NGC 7354      	&107.8$+$02.3		&Emp/Esaph		&1.60     &0.09        	&3       	&96           	&...           	&...        	&N	&N\\
					Sh 2-71      	&035.9$-$01.1		&Bs/Lsbp			&1.14     &0.30       	&14      	&157         &...           	&...        	&N	&N\\
					Sp 1      		&329.0$+$01.9		&Rr/Rsh			&1.13     &0.20        	&6       	&72          	&O(H)	&(Y)        	&...	&P\\
\enddata

\label{table:cycle14}

\tablenotetext{a}{PN and central star data compiled from \cite[and references therein]{Frew:2008ab} unless otherwise noted.}
				
\tablenotetext{b}{Morphologies as listed in Frew (2008, F08): B: bipolar, E: elliptical, R: round, a: asymmetry present, b: bipolar core present, f: filled (amorphous) center, m: multiple shells present, p: point symmetry present, r: ring structure dominant, s: internal structure noted, (h): distinct outer halo. Morphologies following
an abbreviated and very slightly modified version of the classification system described in \cite{Sahai:2011a} (SMV11; see their Table 2): B: bipolar, M:
multipolar, E: elongated, I: irregular, R: round, L: collimated lobe pair, S: spiral arm, c: closed outer lobes, o: open outer lobes; s: CSPN apparent, b: bright
(barrel-shaped) central region, t: bright central toroidal structure; p: point symmetry, a: ansae, i: inner bubble, h: halo; r: radial rays; (/): alternate possibilities.}
				
\tablenotetext{c}{``(Y)" = known binaries with unknown companion types; ``(Y?)" = possible binary; otherwise the type of the companion star is as listed \citep{DeMarco:2009ab}.}
				
\tablenotetext{d}{``Y" = near-IR H$_2$ detected; ``N" = near-IR H$_2$ not detected \cite[][and references therein]{Bernard-Salas:2005ab,Bohigas:2001ab,Webster:1988ab,Ramos-Larios:2013ab,Kastner:1996ab}.}
				
\tablenotetext{e}{X-ray results key: P = point source; D = diffuse source; N = not detected.}

\tablenotetext{f}{\cite{Huckvale:2013ab}}

\tablenotetext{g}{Hillwig et. al. (2014, submitted)}

\tablenotetext{h}{\cite{Lopez:2012ab}}

\tablenotetext{i}{\cite{Herald:2011ab}}

\tablenotetext{j}{\cite{Jones:2014bc}}

\tablenotetext{k}{\cite{Drew:2014a}}

\end{deluxetable}

\begin{deluxetable}{l c c c}
\tabletypesize{\footnotesize}
\tablecolumns{4} 
\tablewidth{0pt} 
\tablecaption{\sc Log of {\em Chandra} Observations}

\tablehead{\colhead{Name}	&
          \colhead{OBSID} &
          \colhead{date} &
          \colhead{exposure}\\
          \colhead{} &
          \colhead{} &
          \colhead{} &
          \colhead{(ks)}}
\startdata
Abell 65         &14583 & 2013-08-22 & 29.7 \\
   HaWe 13         &14578 & 2013-07-15 & 18.6 \\
      Hb 5         &14596 & 2013-10-18 & 29.1 \\
    HbDs 1         &14575 & 2013-08-10 & 19.5 \\
   He 2-11         &14581         &2013-08-10 	& 29.6\\
   IC 1295         &14585 & 2013-08-04 & 29.7 \\
   IC 2149         &14593 & 2012-12-30 & 28.7 \\
   IC 4637         &14586         &2014-03-08         &29.6\\
IC 5148/50         &14576 & 2013-04-07 & 19.7 \\
    M 1-26         &14584         &2013-08-10 & 29.6 \\
    M 1-41         &14591 & 2013-03-02 & 29.6 \\
  NGC 1501         &14574 & 2012-12-08 & 19.7 \\
  NGC 2899         &14589 & 2013-05-28 & 29.6 \\
  NGC 3918         &14597 & 2013-07-31 & 30.0 \\
  NGC 6026         &14588 & 2013-06-29 & 29.6 \\
  NGC 6072         &14590 & 2014-01-23 & 29.6 \\
  NGC 6153         &14579 & 2014-01-17 & 28.6 \\
  NGC 6337         &14577 & 2013-08-04 & 19.7 \\
  NGC 6369         &14594 & 2013-02-18 & 29.6 \\
  NGC 6894         &14587 & 2013-12-13 & 29.2 \\
  NGC 7076         &14592 & 2013-03-29 & 29.6 \\
  NGC 7354         &14595 & 2013-12-26 & 29.6 \\
   Sh 2-71         &14582 & 2013-10-07 & 29.6 \\
      Sp 1         &14580 & 2013-01-24 & 29.4 \\
\enddata

\label{table:obs}

\end{deluxetable}

\begin{deluxetable}{l c c c c}
\tabletypesize{\footnotesize}
\tablecolumns{5} 
\tablewidth{0pt} 
\tablecaption{\sc Planetary Nebulae X-ray Point Source Characteristics}
\tablehead{\colhead{Name}	&
          \colhead{$N$\tablenotemark{a}} &
          \colhead{$C$\tablenotemark{b}} &
          \colhead{median $E$\tablenotemark{c}} &
          \colhead{$E$ range\tablenotemark{d}}\\
          \colhead{} &
          \colhead{(photons)} &
          \colhead{(ks$^{-1}$)} &
          \colhead{(keV)} &
          \colhead{(keV)}}
\startdata
Abell 65\tablenotemark{e}                        &1      &$0.02\pm0.05$               &0.96        &0.96-1.09\\
 HaWe 13                       &...            &$<0.21$                &...              &...\\
 Hb 5\tablenotemark{f}                           &15:      &$0.51\pm0.09$:               &0.94:        &0.81-1.03\\
 HbDs 1                         &10      &$0.49\pm0.06$               &0.65        &0.56-0.78\\
 He 2-11                       &...            &$<0.14$                &...              &...\\
 IC 1295                       &...            &$<0.13$                &...              &...\\
 IC 2149                       &...            &$<0.14$                &...              &...\\
 IC 4637                       &...            &$<0.14$                &...              &...\\
 IC 5148/50                    &...            &$<0.20$                &...              &...\\
 M 1-26                        &...            &$<0.15$                &...              &...\\
 M 1-41                        &...            &$<0.14$                &...              &...\\
 NGC 1501\tablenotemark{g}                      &...            &$<0.20$:                &...              &...\\
 NGC 2899                      &...            &$<0.14$                &...              &...\\
 NGC 3918\tablenotemark{g}                       &21:      &$0.71\pm0.06$:               &0.85:        &0.66-1.03\\
 NGC 6026\tablenotemark{e}                        &1      &$0.05\pm0.08$               &1.16        &0.90-2.31\\
 NGC 6072                      &...            &$<0.14$                &...              &...\\
 NGC 6153\tablenotemark{g}                      &...            &$<0.14$:                &...              &...\\
 NGC 6337                       &39      &$1.99\pm0.07$               &1.27        &1.03-1.73\\
 NGC 6369\tablenotemark{g}                       &8:      &$0.27\pm0.07$:               &1.06:        &0.96-2.38\\
 NGC 6894                      &...            &$<0.14$                &...              &...\\
 NGC 7076                      &...            &$<0.14$                &...              &...\\
 NGC 7354                      &...            &$<0.14$                &...              &...\\
 Sh 2-71                       &...            &$<0.14$                &...              &...\\
 Sp 1                           &11      &$0.37\pm0.07$               &0.93        &0.83-1.33\\
\enddata

\label{table:ptsrcs}

\tablenotetext{a}{Number of source photons, after background subtraction.}
\tablenotetext{b}{Source photon count rate.}
\tablenotetext{c}{Median source photon energy.}
\tablenotetext{d}{Source photon energy range (25th through 75th percentiles).}	
\tablenotetext{e}{Uncertain point source detections at the positions of known binaries. See \S 3.3.}
\tablenotetext{f}{There is no point source apparent within Hb 5.}
\tablenotetext{g}{Point source counts, count rate (or upper limit), median energy, and energy ranges are uncertain due to presence of diffuse emission component.}

\end{deluxetable}

\begin{deluxetable}{c c c c}
\tabletypesize{\footnotesize}
\tablecolumns{4} 
\tablewidth{0pt} 
\tablecaption{\sc Planetary Nebulae: {\em Chandra} X-ray Detection Statistics}
\tablehead{\colhead{category\tablenotemark{a}}	&
	\colhead{$N$\tablenotemark{b}}			&
	\colhead{$N_{PX}$\tablenotemark{c}}		&
	\colhead{$N_{DX}$\tablenotemark{c}}		}
\startdata
					Entire sample				&59			&21 (36\%)			&16 (27\%)\\
					\hline
					Round/elliptical, F08			&47			&19 (40\%)			&14 (30\%)\\
					Bipolar, F08				&12			&2 (17\%)				&2 (17\%)\\
					\hline
					Round/elliptical/irregular, SMV11	&37		&14 (38\%)			&9 (24\%)\\
					Bipolar/multipolar, SMV11		&22		&7 (32\%)				&7 (32\%)\\
					\hline
					near-IR H$_2$ not detected	&19			&9 (47\%)				&10 (53\%)\\
					near-IR H$_2$ detected		&18			&3 (17\%)				&4 (22\%)\\
					\hline
					known binary CSPN			&19			&11 (58\%)			&2 (11\%)\\
\enddata

\label{table:stats}

\tablenotetext{a}{Morphologies as listed and defined in column 3 of Table 1 and associated footnotes; CSPN binary detections and H$_2$ detections as listed, respectively, in columns 9 and 10 of Table 1.}	
\tablenotetext{b}{Total number of sample PNe in each category.}
\tablenotetext{c}{Number of PNe in each category displaying point-like and/or diffuse X-ray emission in {\em Chandra} imaging (Paper I).}

\end{deluxetable}

			\setcounter{figure}{0}
			
			\begin{figure*}[t]
				\centering$
				\begin{array}{cc}
					\includegraphics[height=40mm]{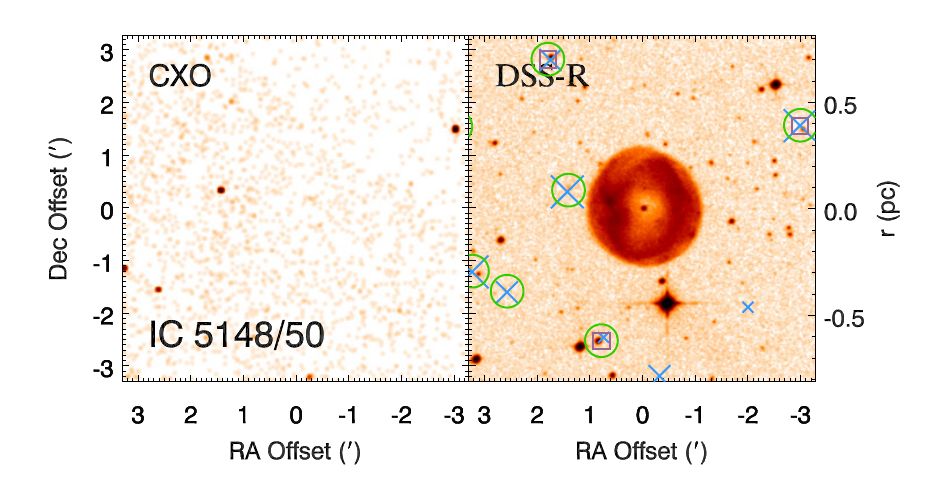}	&\includegraphics[height=40mm]{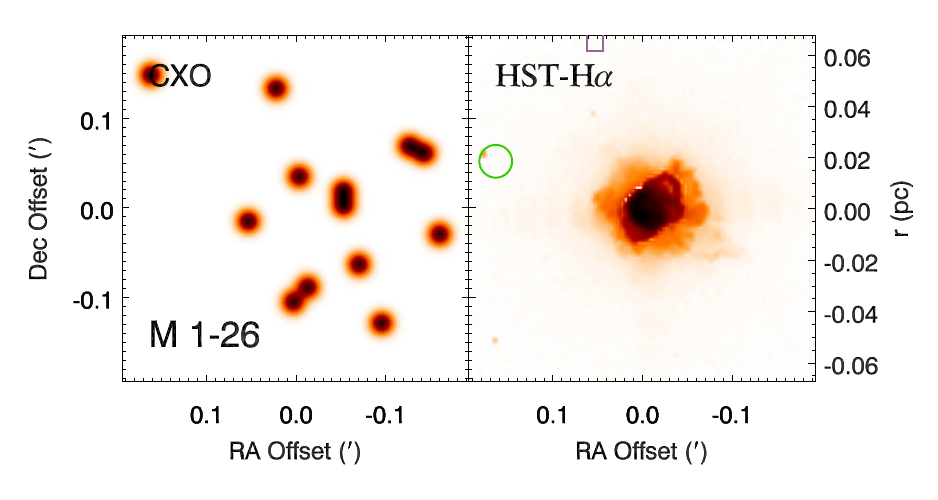}\\
					\includegraphics[height=40mm]{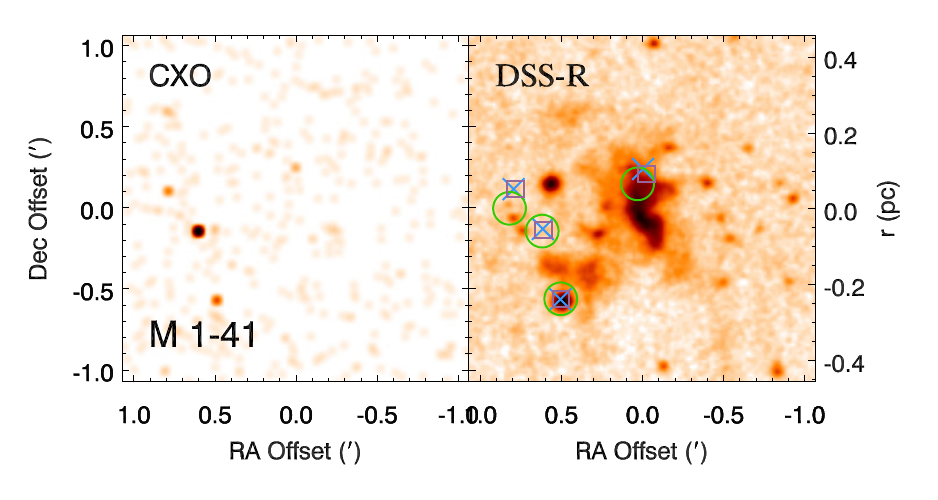}	&\includegraphics[height=40mm]{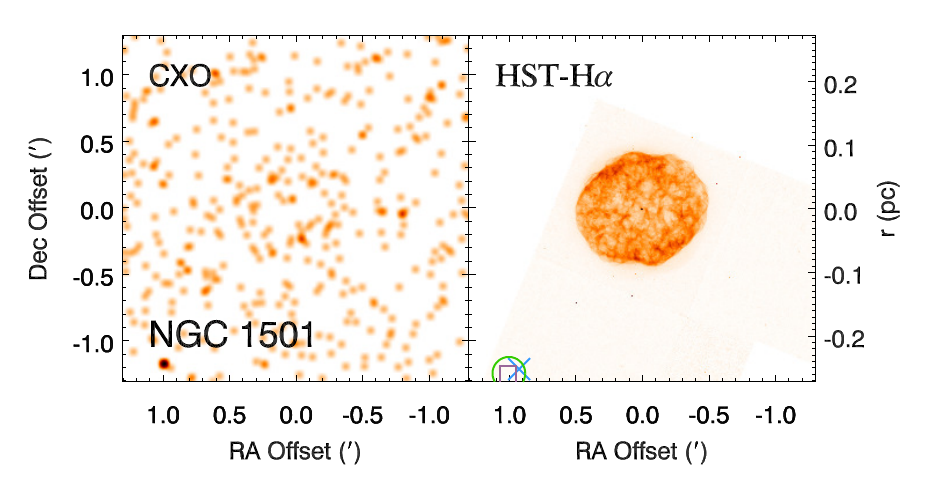}\\
					\includegraphics[height=40mm]{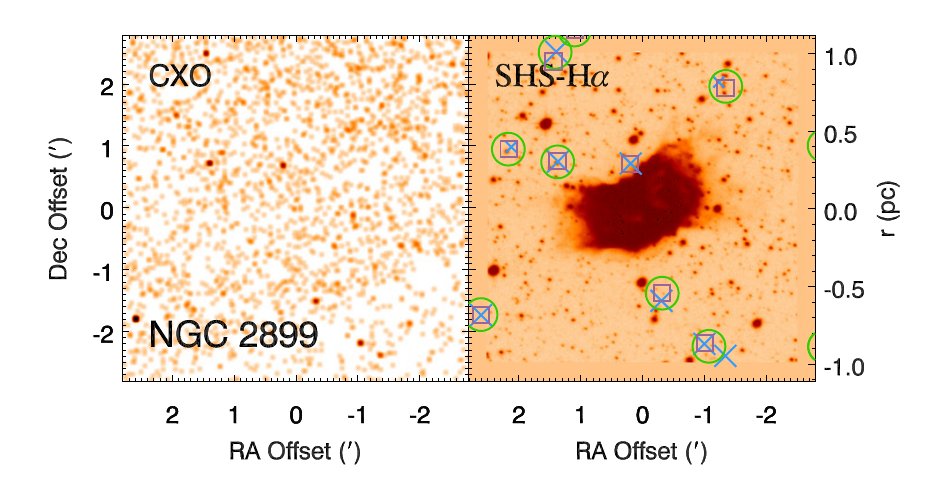}	& \includegraphics[height=40mm]{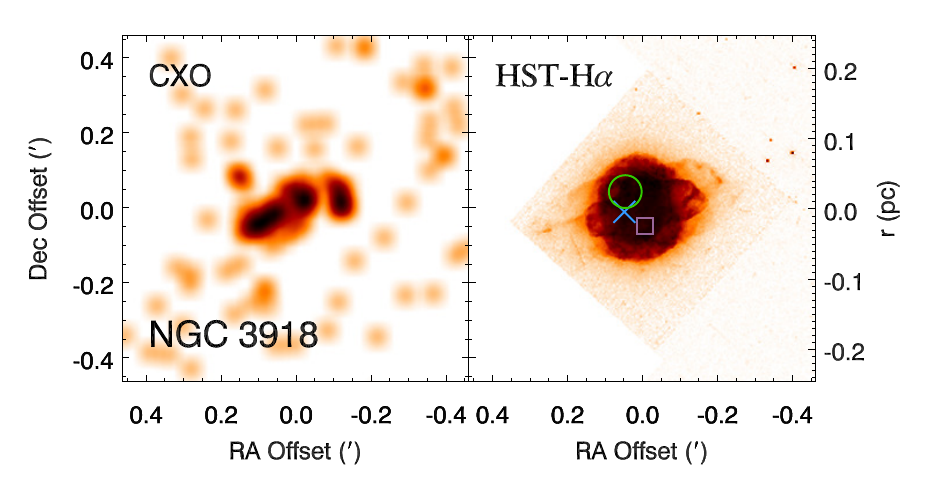}\\
					\includegraphics[height=40mm]{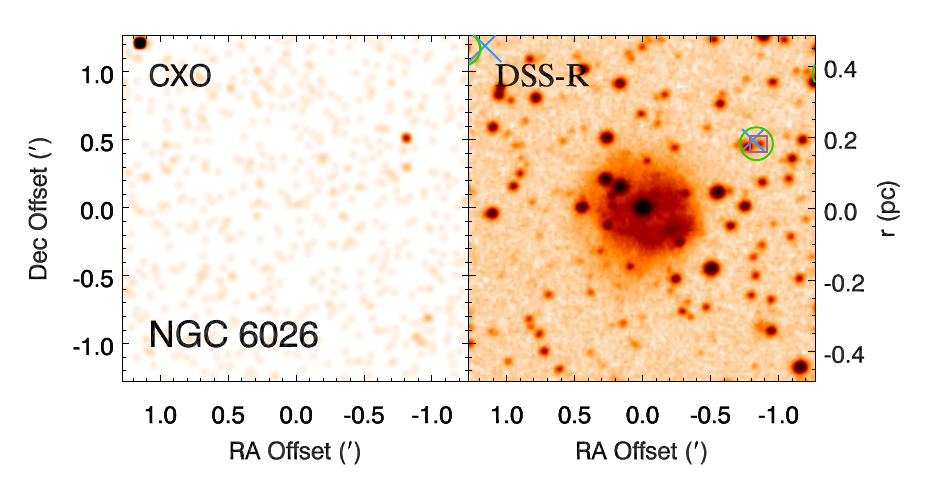}	& \includegraphics[height=40mm]{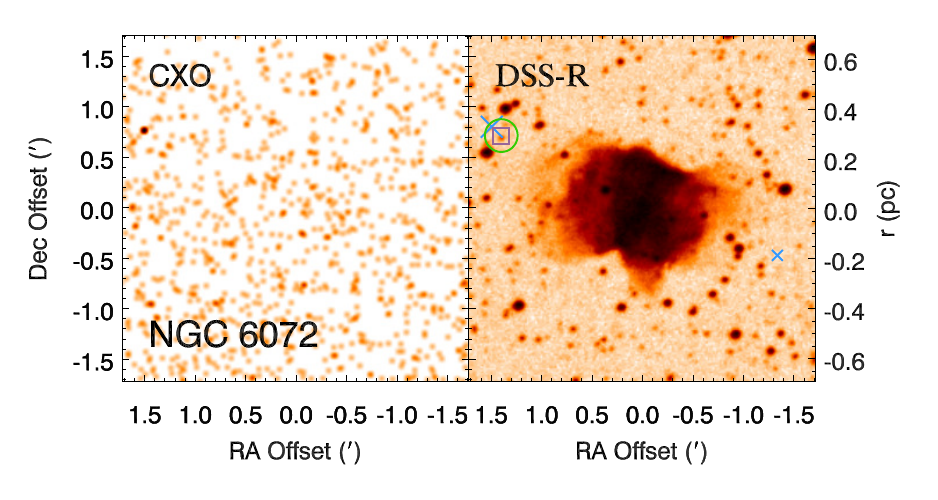}
				\end{array}$
				\caption{\label{fig:images2}\footnotesize{(cont.)}}
			\end{figure*}
			
			\setcounter{figure}{0}
			
			\begin{figure*}[t]
				\centering$
				\begin{array}{cc}
					\includegraphics[height=40mm]{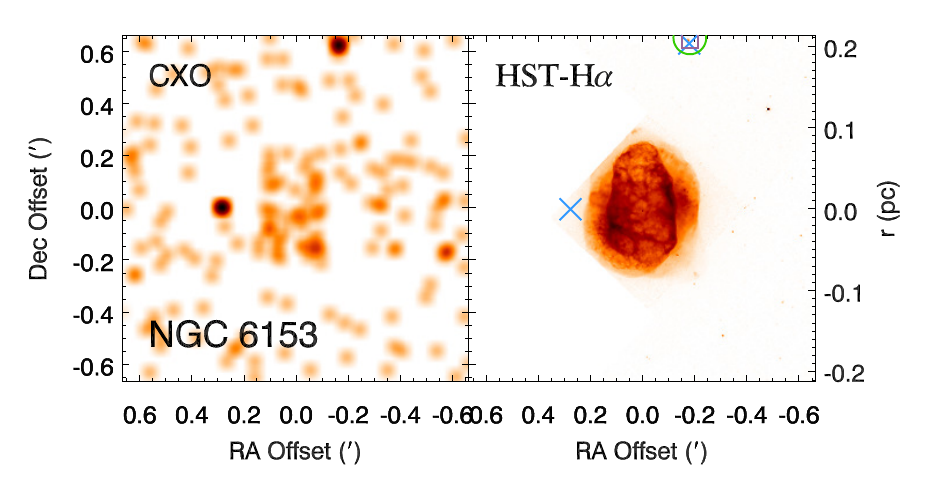}	& \includegraphics[height=40mm]{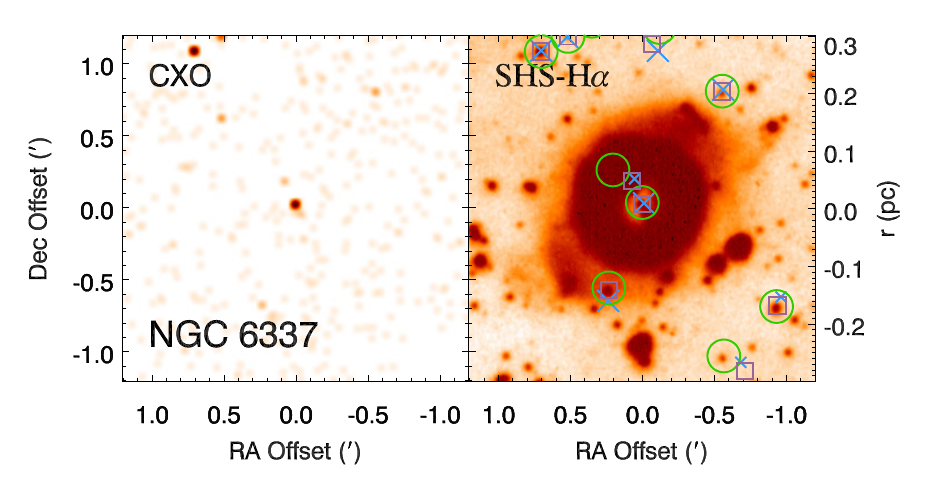}\\
					\includegraphics[height=40mm]{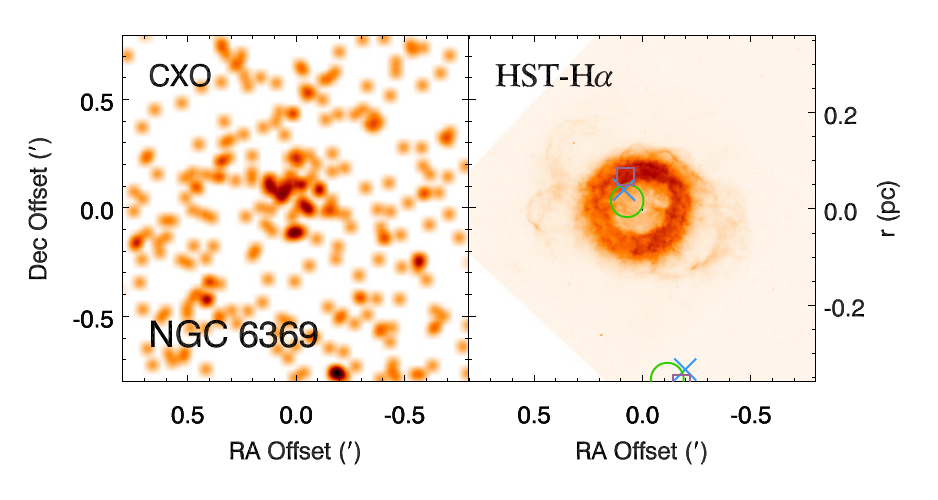}	&\includegraphics[height=40mm]{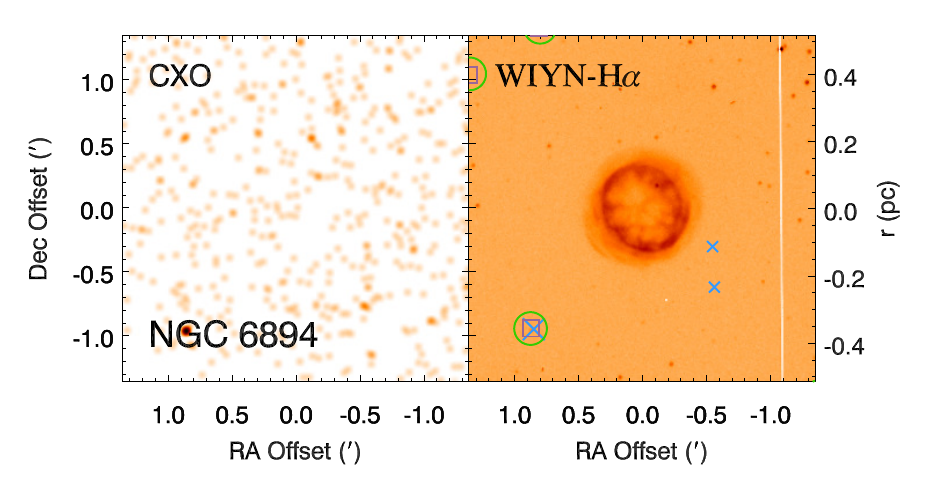}\\
					\includegraphics[height=40mm]{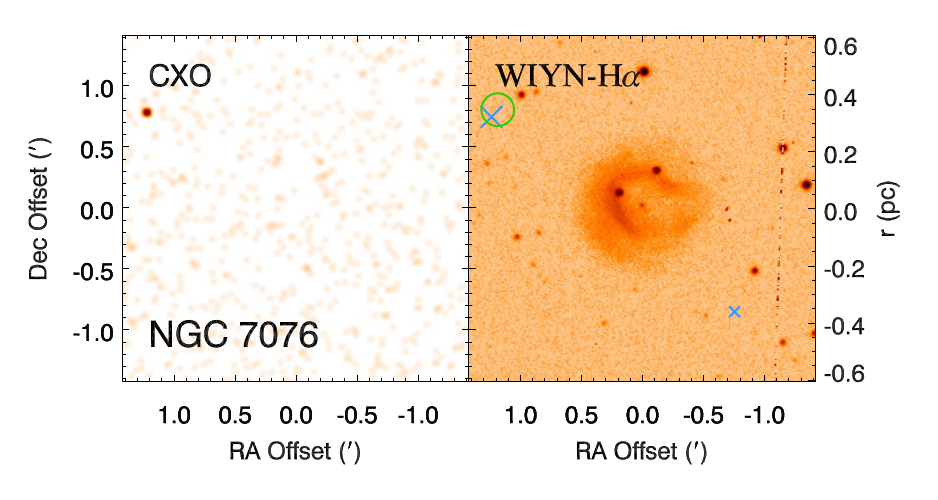}	&\includegraphics[height=40mm]{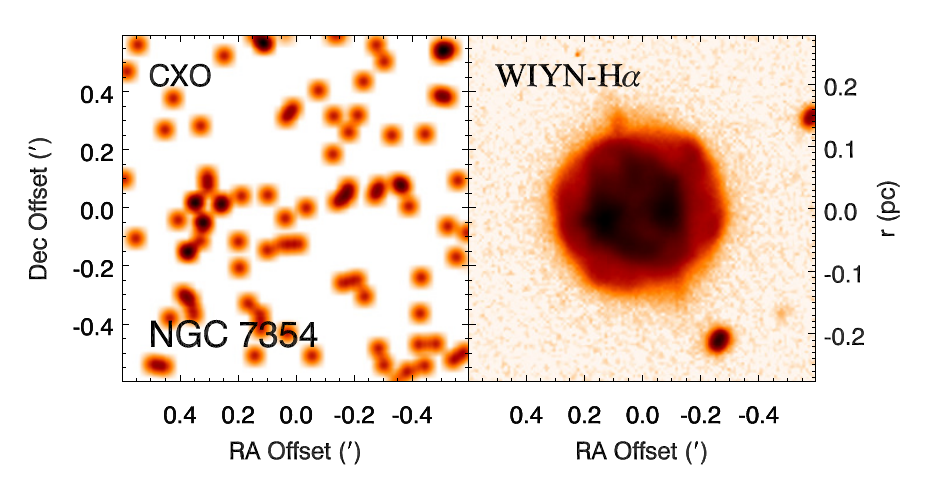}\\
					\includegraphics[height=40mm]{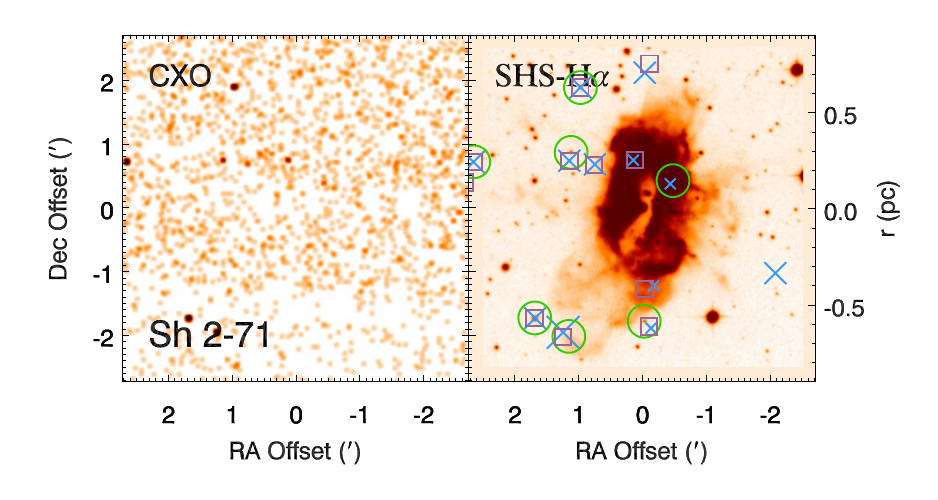}	& \includegraphics[height=40mm]{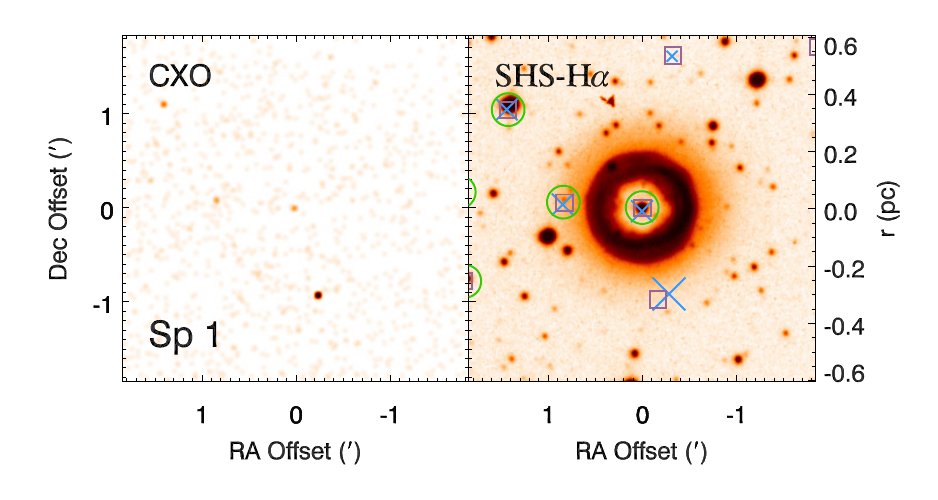}
				\end{array}$
				\caption{\label{fig:images3}\footnotesize{(cont.)}}
			\end{figure*}
			
			\begin{figure}[t]
				\centering
					\includegraphics[width=150mm]{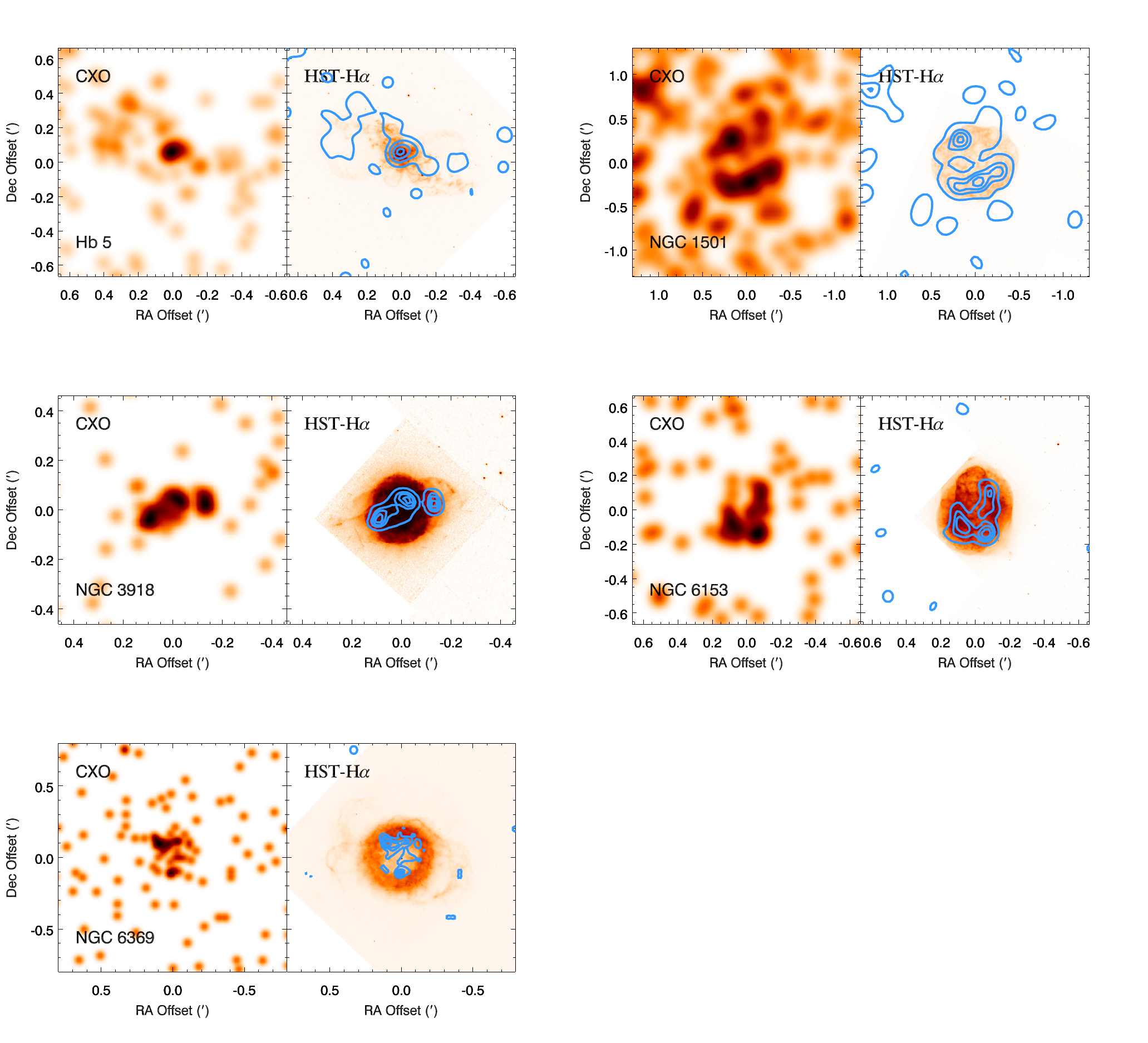}
				\caption{\label{fig:diff}\footnotesize{Images of Table 1  PNe in which diffuse X-ray emission has been detected by {\em Chandra}. The left and right panel for each PN display, respectively, {\em Chandra} 0.5-1.2 keV images filtered to efficiently remove the soft background and {\em Chandra} contours overlaid on optical images. The {\em Chandra} images of Hb 5, NGC 1501, and NGC 6153 have been smoothed with a $12.5''$ FWHM Gaussian, while the {\em Chandra} images of NGC 3918 and NGC 6369 have been smoothed with a FWHM of $3''$. The contour levels of all but Hb 5 are 30, 60, 75, and 90\% of the maximum X-ray surface brightness (Hb 5 has contour levels of 10, 30, 60, and 90\%).}}
			\end{figure}
					
			\begin{figure}[t]
				\centering
					\includegraphics[width=80mm]{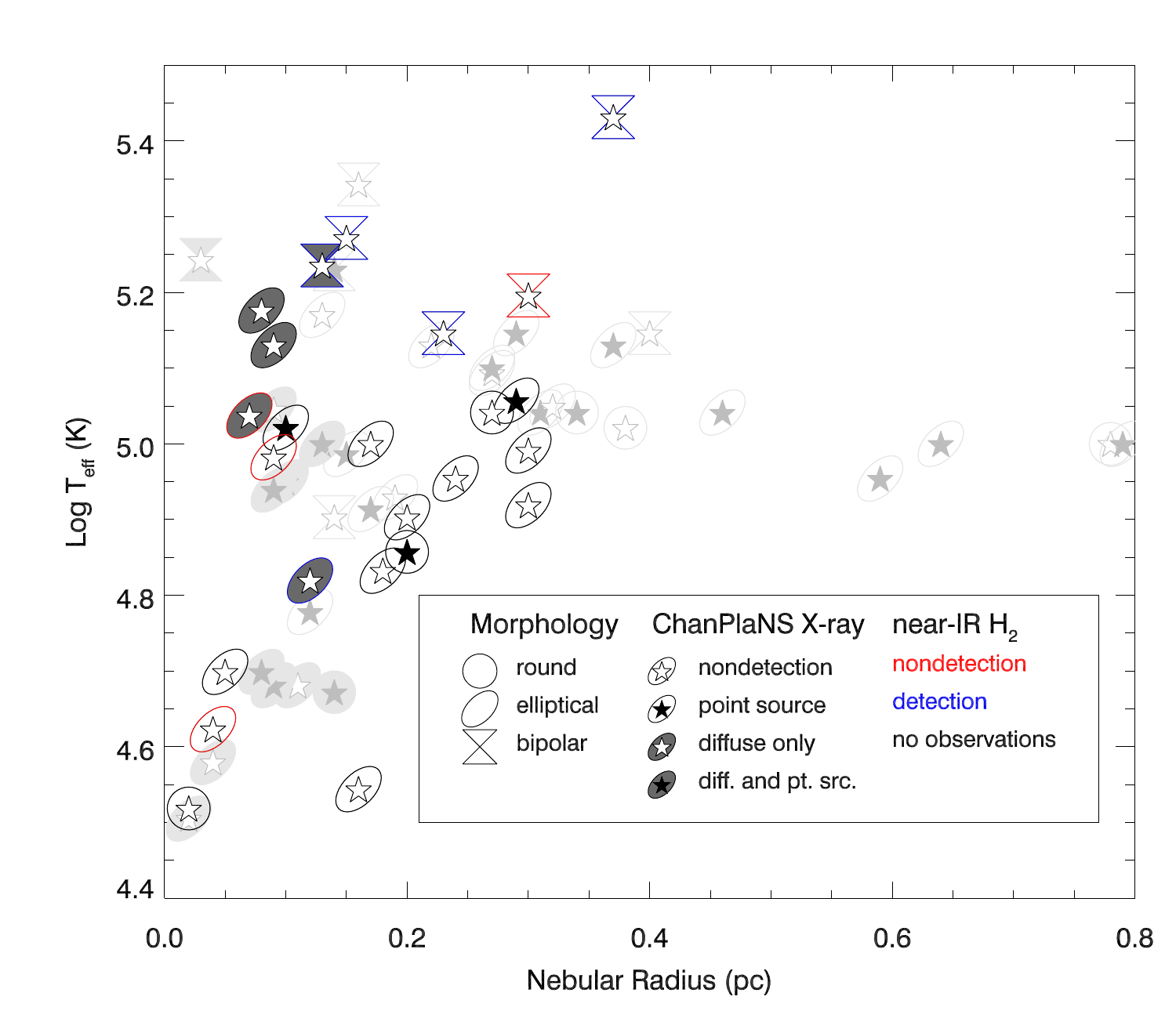}
					\includegraphics[width=80mm]{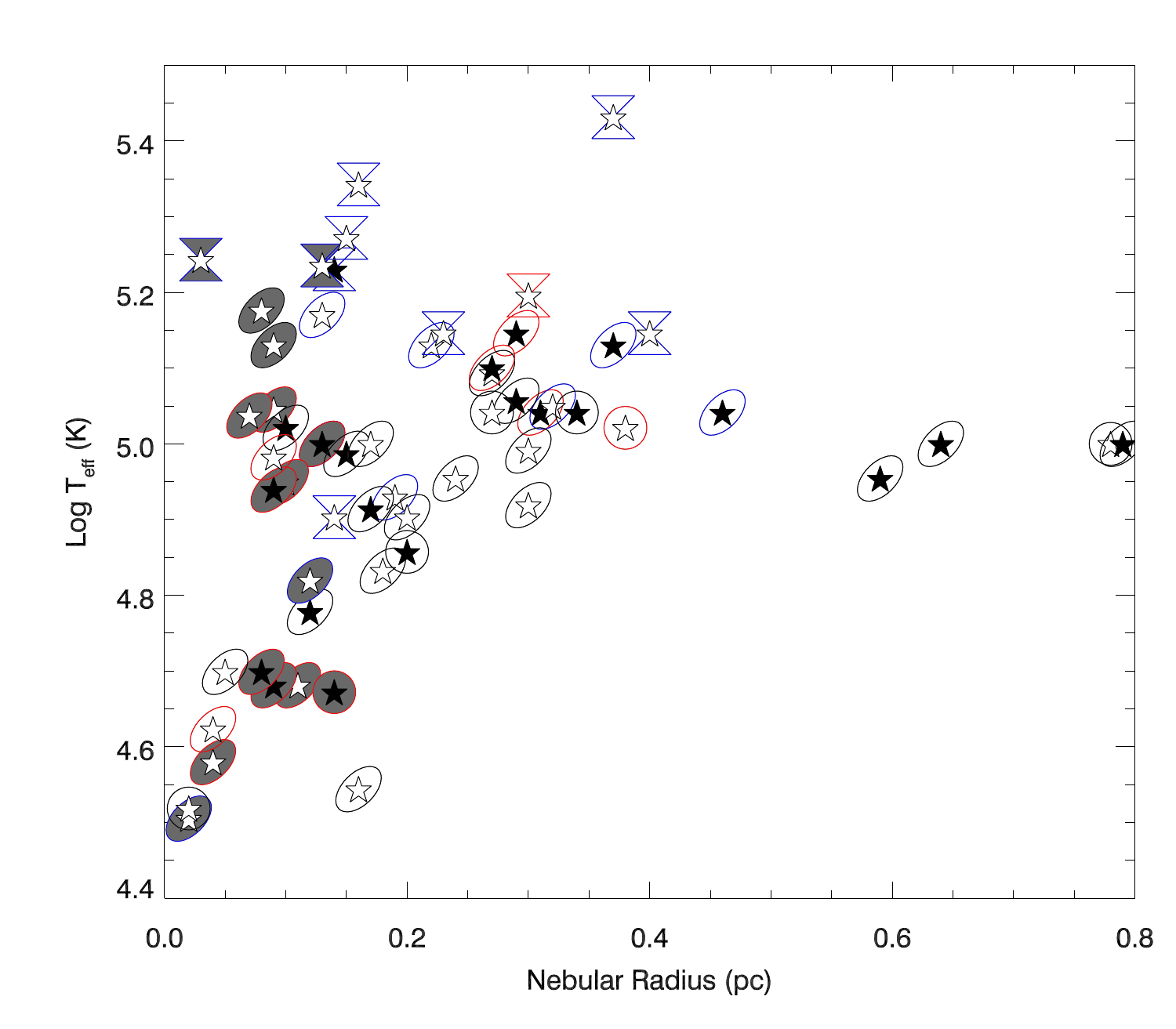}
				\caption{\label{fig:morph}\footnotesize{Plots of CSPN $T_{\text{eff}}$ vs.~PN radius for all ChanPlaNS observed objects, with symbols indicating presence or absence of diffuse or point-like X-ray emission, as well as PN morphology and presence or absence of H$_2$ emission. In the left panel, the highlighted objects represent Cycle 14 data and the light gray objects represent Cycle 12 and archival data (Paper I).}}
			\end{figure}
			
			\begin{figure}[t]
				\centering
					\includegraphics[width=100mm]{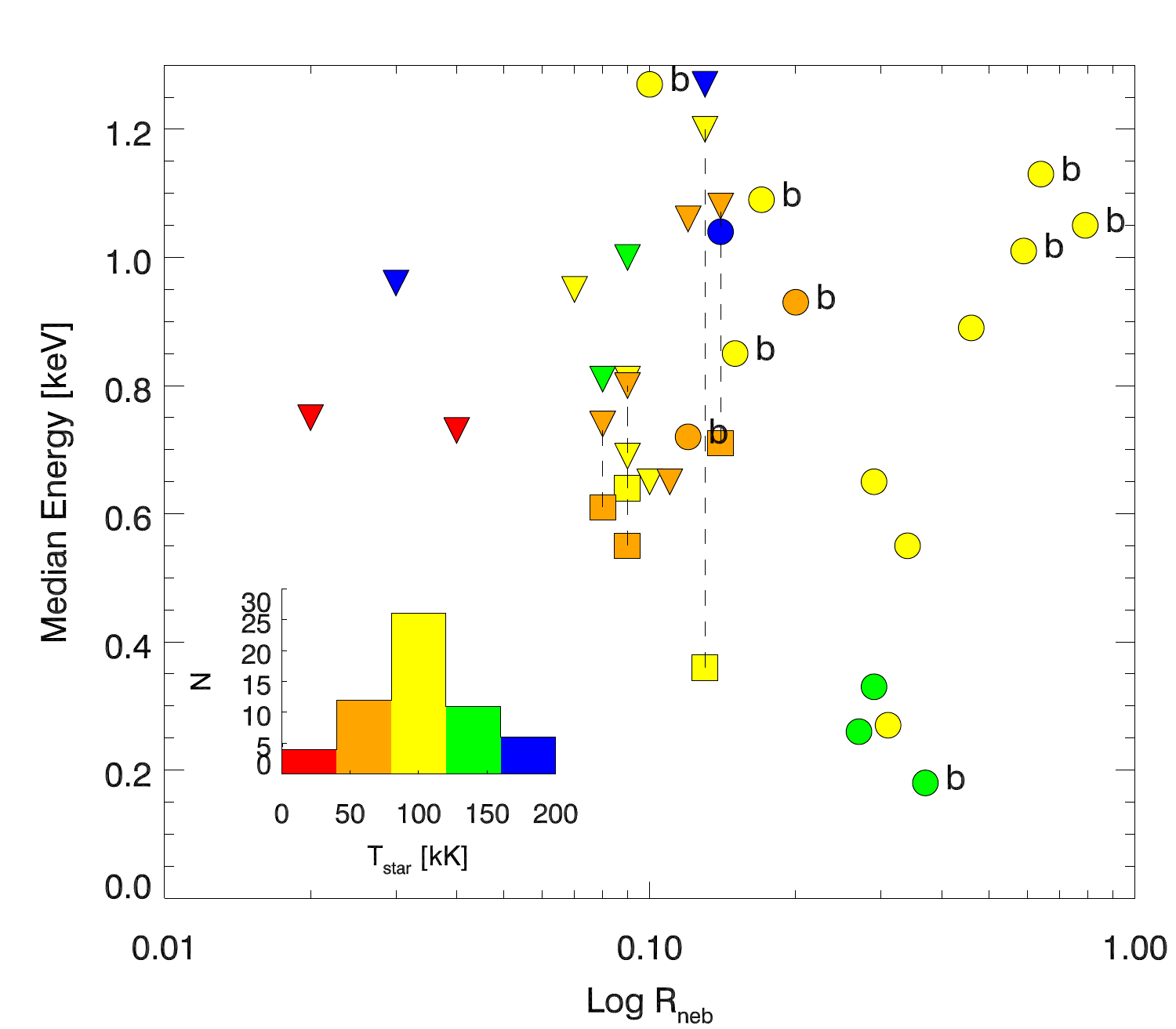}
				\caption{\label{fig:medenergy}\footnotesize{Median X-ray photon energy vs.~nebular radius for all Cycle 12+14+archival sample PNe detected as X-ray sources. Symbol shapes indicate the nature of the PN X-ray source (triangles: diffuse sources; circles: CSPN point sources; squares: diffuse+CSPN sources) and are color-coded according to CSPN $T_{\text{eff}}$, whose distribution for the observed sample is shown in the inset histogram. Known binary CSPNe are indicated by ``b". The dashed line connects the diffuse+CSPN emission with the corresponding diffuse emission for sources that exhibit both types of emission.}}
			\end{figure}
			
			\begin{figure}[t]
				\centering
					\includegraphics[width=80mm]{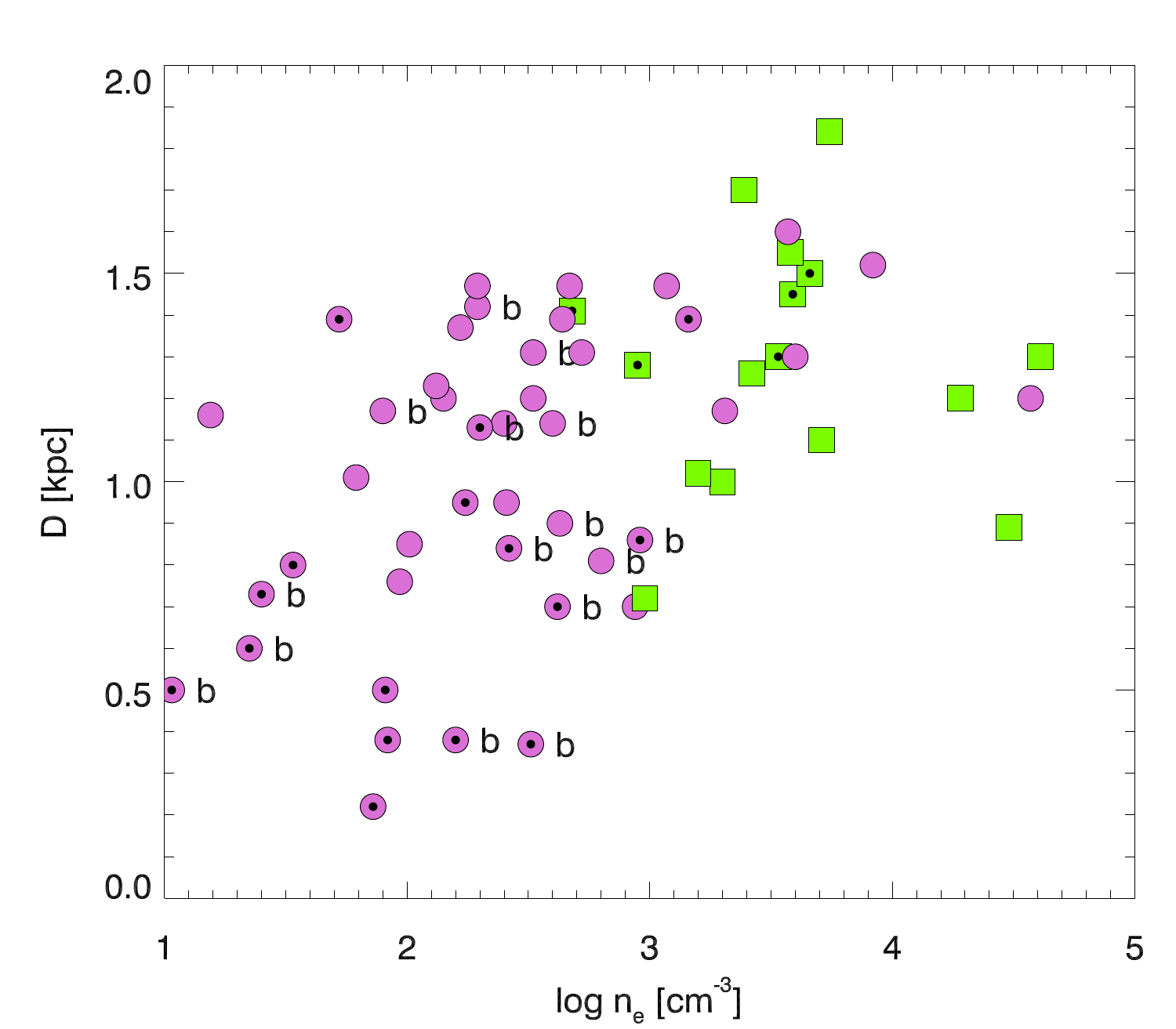}
					\includegraphics[width=80mm]{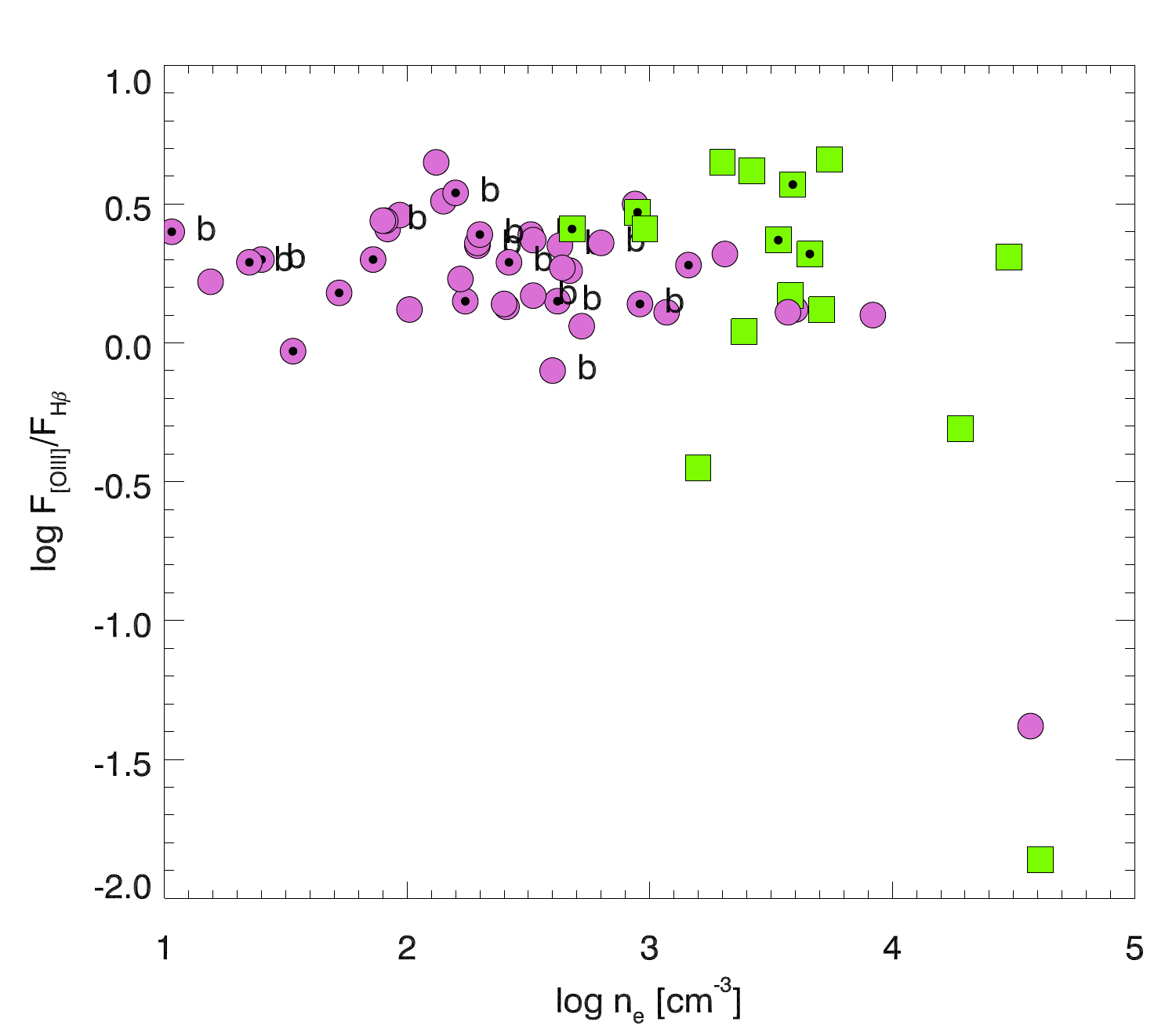}
				\caption{\label{fig:ne}\footnotesize{Distance ({\em left}) and ratio of [\ion{O}{3}] to H$\beta$ fluxes ({\em right}) vs.~the log of electron density $n_e$ for PNe detected (green squares) and not detected (purple circles) as \chanplans\ diffuse X-ray sources. A bullet within each symbol indicates that the CSPNe was detected as a point-like X-ray source. A ``b" next to a symbol indicates the CSPN is a known binary.}}
			\end{figure}

\end{document}